\documentclass[12pt,letterpaper]{article}
\usepackage[utf8]{inputenc}

\pdfoutput=1
\usepackage{color}
\definecolor{darkblue}{rgb}{0.1,0.1,.7}
\usepackage[colorlinks, linkcolor=darkblue, citecolor=darkblue, urlcolor=darkblue, linktocpage]{hyperref}
\usepackage[]{amsmath}
\usepackage{amssymb}
\usepackage[]{graphicx}
\usepackage[]{latexsym}
\usepackage[utf8]{inputenc}
\usepackage{slashed,graphicx,color,amsmath,amssymb}
\usepackage[mathscr]{eucal}
\usepackage{mathrsfs}
\usepackage[margin=10pt,font=small,labelfont=bf]{caption}
\usepackage{amscd}
\usepackage{bm}
\usepackage{xcolor}
\usepackage{upgreek}
\usepackage[square, comma, sort&compress,numbers]{natbib}
\usepackage[all,cmtip]{xy}
\usepackage[margin=1.5in]{geometry}
\usepackage{cleveref}
\usepackage[color=cyan!30!white,linecolor=red,textsize=footnotesize]{todonotes}
\usepackage{microtype}
\usepackage{soul}
\numberwithin{equation}{section}
\numberwithin{figure}{section}

\hypersetup{
	pdfstartview={FitH},    
	pdftitle={Operator growth in 2d CFT},    
	pdfauthor={Caputa, Datta},     
	colorlinks=true,       
	linkcolor=blue,          
	citecolor=blue!59!black! ,        
	filecolor=magenta,      
	urlcolor=blue           
}

\def\pd{\partial}

\def\cN{\mathcal{N}}

\def\ie{{\it i.e.~}}
\def\eg{{\it e.g.~}}

\def\nn{\nonumber}
\def\pd{\partial}

\def\l1{{{1-loop}}}

\def\n1{\Bigg|_{n=1}}

\def\n{{(n)}}

\def\O{\mathcal{O}}
\def\cN{\mathcal{N}}

\def\cL{\mathcal{L}}

\def\pd{\partial}

\def\beq{\begin{equation}}
\def\eeq{\end{equation}}

\def\bea{\begin{eqnarray}}
\def\eea{\end{eqnarray}}
\def\nn{\nonumber}
\def\pd{\partial}

\def\l1{{\text{1-loop}}}

\def\n1{\Bigg|_{n=1}}

\def\n{{(n)}}

\def\O{\mathcal{O}}
\def\cN{\mathcal{N}}

\def\bra#1{{\langle}#1|}

\def\ket#1{|#1\rangle}

\def\vev#1{\langle{#1}\rangle}

\def\be{\begin{equation}}
\def\ee{\end{equation}}
\def\bal{\begin{array}{l}}
\def\ba#1{\begin{array}{#1}}  
	\def\ea{\end{array}}
\def\bea{\begin{eqnarray}}
\def\eea{\end{eqnarray}}
\def\beas{\begin{eqnarray*}}
	\def\eeas{\end{eqnarray*}}

\def\nn{\\\nonumber}

\def\vev#1{\langle #1 \rangle}

\def\nn{\nonumber}
\def\bit{\begin{item}}
	\def\eit{\end{item}}
\def\benu{\begin{enumerate}}
	\def\eenu{\end{enumerate}}

\def\cO{{\mathcal O}}

\def\p{\partial}

\def\cW{\mathcal{W}}

\DeclareFontFamily{U}{wncy}{}
\DeclareFontShape{U}{wncy}{m}{n}{<->wncyr10}{}
\DeclareSymbolFont{mcy}{U}{wncy}{m}{n}
\DeclareMathSymbol{\Sha}{\mathord}{mcy}{"58}

\newdimen\tableauside\tableauside=1.0ex
\newdimen\tableaurule\tableaurule=0.4pt
\newdimen\tableaustep
\def\phantomhrule#1{\hbox{\vbox to0pt{\hrule height\tableaurule width#1\vss}}}
\def\phantomvrule#1{\vbox{\hbox to0pt{\vrule width\tableaurule height#1\hss}}}
\def\sqr{\vbox{%
		\phantomhrule\tableaustep
		\hbox{\phantomvrule\tableaustep\kern\tableaustep\phantomvrule\tableaustep}%
		\hbox{\vbox{\phantomhrule\tableauside}\kern-\tableaurule}}}
\def\squares#1{\hbox{\count0=#1\noindent\loop\sqr
		\advance\count0 by-1 \ifnum\count0>0\repeat}}
\def\tableau#1{\vcenter{\offinterlineskip
		\tableaustep=\tableauside\advance\tableaustep by-\tableaurule
		\kern\normallineskip\hbox
		{\kern\normallineskip\vbox
			{\gettableau#1 0 }%
			\kern\normallineskip\kern\tableaurule}%
		\kern\normallineskip\kern\tableaurule}}
\def\gettableau#1 {\ifnum#1=0\let\next=\null\else
	\squares{#1}\let\next=\gettableau\fi\next}

 \makeatletter
 \g@addto@macro\bfseries{\boldmath}
 \makeatother

\makeatletter
\if@todonotes@disabled

\else

\fi
\makeatother

\begin{document}

\definecolor{tinge}{RGB}{255, 244, 195}
\sethlcolor{tinge}
\setstcolor{red}

\vspace*{-.8in} \thispagestyle{empty}
\begin{flushright}
	\texttt{CERN-TH-2021-151}
\end{flushright}
\vspace{.5in} {\Large
\begin{center}
{\LARGE \bf  Operator growth in 2d CFT}
\end{center}}
\vspace{.4in}
\begin{center}
{Pawel Caputa$^1$ \& Shouvik Datta$^2$}
\\
\vspace{.4in}
\small{
	$^1$\textit{Faculty of Physics, University of Warsaw,\\ 
		ul.~Pasteura 5, 02-093 Warsaw, Poland. }\\
	\vspace{.5cm}
  $^2$\textit{Department of Theoretical Physics, CERN,\\
	1 Esplanade des Particules, Geneva 23, CH-1211, Switzerland.}\\
\vspace{.5cm}

} \vspace{0cm}
\begingroup\ttfamily\small
pawel.caputa@fuw.edu.pl,~sdatta@cern.ch\par
\endgroup


\end{center}

\vspace{.6in}

\begin{abstract}
\normalsize We investigate and characterize the dynamics of operator growth in  irrational two-dimensional conformal field theories. By employing the oscillator realization of the Virasoro algebra and CFT states, we systematically implement the Lanczos algorithm and evaluate the Krylov complexity of simple operators (primaries and the stress tensor) under a unitary evolution protocol. Evolution of primary operators proceeds as a flow into the `bath of descendants' of the Verma module. These descendants are labeled by integer partitions and have a one-to-one map to Young diagrams. This relationship allows us to rigorously formulate operator growth as paths spreading along the Young's lattice. We extract quantitative features of these paths and also identify the one that saturates the conjectured upper bound on operator growth. 
\end{abstract}

\vskip 0.7cm \hspace{0.7cm}

\newpage

\setcounter{page}{1}

\noindent\rule{\textwidth}{.1pt}\vspace{-1.2cm}
\begingroup
\hypersetup{linkcolor=black}
\tableofcontents
\endgroup
\noindent\rule{\textwidth}{.2pt}

\setlength{\abovedisplayskip}{15pt}
\setlength{\belowdisplayskip}{15pt}

\section{Introduction}
\label{sec:intro}
Quantum systems approach thermal equilibrium despite undergoing unitary evolution. A route to understand this mechanism is to analyze the spread of quantum information as the system evolves. Information initially stored in simple, local operators scrambles into a large number of degrees of freedom at later times. A precise formulation of this notion of operator growth is key to understanding the emergence of irreversible macroscopic behaviour, such as hydrodynamics or thermalization  \cite{deutsch1991quantum,srednicki1994chaos,vonKeyserlingk:2017dyr,Khemani:2017nda}. 

There are several diagnostics for operator growth in quantum chaotic systems. The key idea is to analyze the Heisenberg evolution of a local operator, $\O(t)= e^{iHt}\O(0)e^{-iHt}$, and observe the effects it has on the system at late times. A simple application of the Baker-Campbell-Hausdorff identity gives a sum over increasingly nested commutators of the Hamiltonian
\begin{align}
	e^{iHt}\O(0)e^{-iHt} = \O(0)  + it[H,\O(0)] - \frac{t^2}{2}[H,[H,\O(0)]] - \frac{it^3}{6}[H,[H,[H,\O(0)]] ] + \cdots ~. \nn 
\end{align}
This shows that even if $\O$ is a simple local operator and the Hamiltonian has few-body interactions, the effect of the operator spreads all throughout the system as time evolves. One way this phenomenon manifests itself is by the inability of $\O(t)$ to commute with other simple probe operators; this is captured by  the out-of-time ordered correlators (OTOCs) \cite{Shenker:2013pqa,KitaevTalk,Maldacena:2015waa}. 

It has recently emerged that a more direct means to characterize the growth of the operator $\O$ can be achieved by using the recursion method \cite{Viswanath1990}. In this framework, the nested commutators above are defined as operators obtained by applying the Liouvillian, $\cL \equiv [H,\ast  ]$, on the operator $\O$. This is the action,  $\tilde{\O}_n=\cL^n \O$. After introducing an inner product in the space of operators, one can use the Lanczos algorithm  to build an orthonormal basis associated with the set  $\tilde{\O}_n$ \cite{Lanczos:1950zz}. This basis is known as  the Krylov basis. In addition to the basis, the  algorithm yields a set of coefficients, the so-called Lanczos coefficients, that encode transition amplitudes between two orthonormal operators. 
These coefficients have specific growth properties depending on whether the system is non-interacting, integrable or chaotic. As with the chaos bound for OTOCs \cite{Maldacena:2015waa,Murthy:2019fgs}, the maximal growth of Lanczos coefficients, is conjectured to be linear \cite{Parker:2018yvk}.  

In the Krylov basis, the Liouvillian acquires a tridiagonal matrix representation, with the matrix elements being nothing but the Lanczos coefficients or the transition amplitudes. Information of the coefficients also precisely underpins the rate of delocalization of the operator at late times. This is encapsulated through a measure known as Krylov complexity (or K-complexity). For chaotic systems with maximal Lanczos coefficients (such as the SYK model \cite{Sachdev:1992fk}), this quantity shows an exponential growth. This confirms the physical expectation that simple operators irreversibly grow into those with higher `complexity'. In contrast to OTOCs or the Eigenstate Thermalization Hypothesis (ETH), the Liouvillian approach transcends the need of additional probe operators. As a result, one can isolate the truly universal aspects of quantum information spreading.

\begin{figure}[t!]
	\centering
	\includegraphics[width=0.97\linewidth]{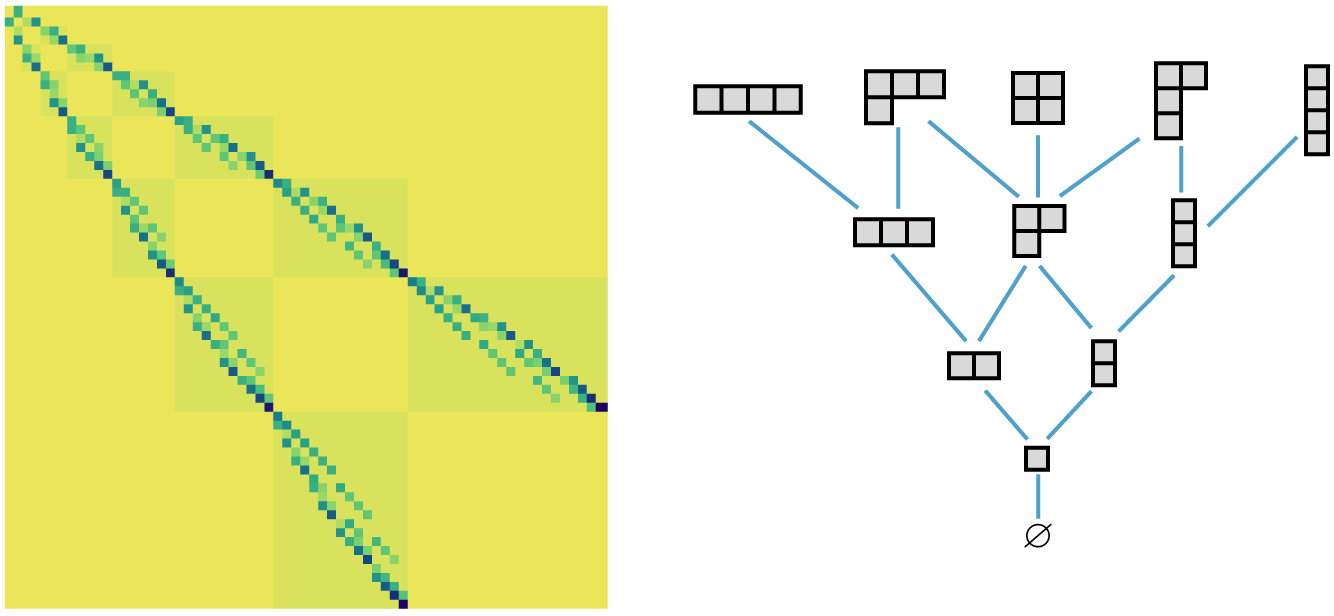}
	\caption{[Left] Block tridiagonal structure of the Liouvillian matrix: the matrix elements are indexed by orthogonal descendants. Each block specifies transition amplitudes between two descendant states of adjacent levels and the individual values are the Lanczos coefficients. [Right] The Young's lattice: the vertices correspond to integer partitions or descendant states and the edges are weighted by the Lanczos coefficients.}
	\label{fig:intro-fig}
\end{figure}
In this work, our goal is to utilize the above paradigm to investigate the  details of operator growth in two-diemnsional conformal field theories.  Motivated by semiclassical holographic duals to AdS$_3$ gravity, we shall focus on irrational CFTs with Virasoro symmetry. These CFTs are expected to the thermalize owing to presence of black holes in the gravity dual. However, there are also an infinite number of conserved quantities (KdV charges) which imply that the standard mechanism postulated by ETH require refinements \cite{Datta:2019jeo,Besken:2019bsu,Dymarsky:2019etq}. Along the same lines, due to additional symmetries there are exact degeneracies present in the spectrum and, therefore, we require a systematic reorganization of the Lanczos coefficients that will arise on repeated applications of the Liouvillian.  We undertake this task by utilizing the oscillator formulation of the Virasoro algebra and its representations \cite{zamolodchikov1986two}. Amongst some other advantages, this formalism provides a convenient, orthogonal basis that can be efficiently used to track the time evolution of simple operators, such as Virasoro primaries and the stress tensor. Unlike the conventional Lanczos algorithm for quantum systems without degeneracies, it is not possible to shrink all the information of $\cL^n O$ into a single state/operator in 2d CFT. 
The matrix representation of the Liouvillian  generalizes and it acquires a block-tridiagonal structure. We analytically evaluate the matrix elements, \ie the Lanczos coefficients, and study their properties -- a matrix plot of this is shown in \cref{fig:intro-fig} (left). Moreover, the oscillator formalism also allows us to obtain a closed form for the time-dependent state/wavefunction under the Liouvillian evolution. The K-complexity extracted from this wavefunction shows the expected exponential behaviour for operator growth. This quantity can also be interpreted as the volume in the information geometry, given by the Fubini-Study metric which is a well-known distance measure between quantum states.  We characterize the evolution further by evaluating the Renyi entropies.

Our findings are fixed purely by Virasoro symmetry. The main picture that emerges from our analysis is that the growth of a primary operator can be elegantly described  as spreading on a network or graph, known as the Young's lattice -- see \cref{fig:intro-fig} (right). When we start out with a single primary operator, its evolution proceeds as a decay of its wavefunction into the set of descendants of its Verma module -- in a sense, this is propagation in momentum space into states with increasingly higher energies. The descendants of the primary operator are, in turn, labeled by integer partitions of the descendant level and, therefore, these have a one-to-one correspondence with Young diagrams. The specific evolution protocol we work with corresponds to adding or removing a box from the Young diagram. The descendants or the Young diagrams are the vertices or nodes of the Young's lattice, whilst the Lanczos coefficients are weights of the edges or links of the lattice. This graphical correspondence provides a valuable context to organize the structural properties of operator growth for 2d CFTs.\footnote{See \cite{Bentsen:2018uph,Chen:2019klo,Kim:2021okd} for recent progress on understanding operator growth from a graph theory perspective.}  The late-time behaviour maps to the asymptotic regime of high descendant levels. We study the evolution along  paths in the lattice (both typical and atypical ones) and also identify the path that saturates the conjectured upper bound on the growth of Lanczos coefficients.  Moreover, using some well-established results from combinatorics and asymptotics of Young diagrams we provide bounds, tighter than those recently conjectured in \cite{Avdoshkin:2019trj}, on the number of paths leading to a specific descendant at late times. 

This paper is structured as follows. In \cref{sec:prelims} we introduce the notions of Liouvillian evolution, Lanczos coefficients, K-complexity and specify our evolution protocol. 
This section also contains a brief review of the main workhorse for our computations, the oscillator formalism of Virasoro algebra. We also spell out the relation of the descendants to Young diagrams and introduce Young's lattice in this section. In \cref{sec:liov} we evaluate the Liouvillian evolution, find the Lanczos coefficients and describe the operator growth for primaries through the Young's lattice. Section \ref{sec:evo} provides an exact expression for the evolved state from which we evaluate the K-complexity, its fluctuations and Renyi entropies. We analyze the operator growth of the stress tensor by calculating the associated K-complexity in \cref{sec:stress}. We conclude and discuss generalizations  in \cref{sec:discuss}. The appendices \ref{sec:master-id} and \ref{app:norms} contain proofs of some crucial identities and some consistency checks. 


\section{The ingredients}
\label{sec:prelims}
\def\Op{\mathcal{O}}
\subsection{Notions of operator growth}
\label{subsec:complex-defs}

Characterizing operator growth in quantum systems is in general a complicated problem. The main reason is that the notion of the operator size, or operator complexity, may not be a universally well-defined concept and one has to work with specific model-dependent tools that probe the growth. This subject has been actively growing in many-body systems \cite{Parker:2018yvk,vonKeyserlingk:2017dyr,Jian:2020qpp,Kudler-Flam:2020yml,MacCormack:2020auw,Mascot:2020qep}  with an increasing interest from holography \cite{Brown:2015lvg,Magan:2018nmu,Roberts:2018mnp,Susskind:2020gnl,Qi:2018bje,Lin:2019qwu,Haehl:2021prg,Haehl:2021dto}.   

An interesting progress towards a universal measure of the operator complexity has been made in \cite{Parker:2018yvk}. One starts with the Heisenberg operator $\Op(t)$ that can be formally expanded in a series of nested commutators with the Hamiltonian
\be\label{Heisenberg}
\Op(t)=e^{iHt}\,\Op(0)\, e^{-iHt}=\sum^\infty_{n=0}\frac{(it)^n}{n!}\tilde{\Op}_n~,
\ee
such that operators on the right are 
\be
\tilde{\Op}_0=\Op,\quad \tilde{\Op}_1=[H,\Op],\quad \tilde{\Op}_2=[H,[H,\Op]],\quad \cdots~. 
\ee
As time progresses, a simple operator $\Op(0)$ ``grows" in the space of operators of the theory becoming more ``complex". The goal is to quantify this growth in a precise manner. 

We describe the standard procedure for arbitrary quantum systems first and then, in the next section, we will see the explicit realization for 2d CFTs. With these objectives in mind, we  introduce a Liouvillian super-operator 
\begin{align}
	\mathcal{L}=[H,\ast]~,
\end{align}
such that the nested commutators $\tilde{\Op}_n$ come from its recursive action on $\Op(0)$, \ie  $\tilde{\Op}_n=\mathcal{L}^n\Op(0)$. The Heisenberg evolution \eqref{Heisenberg} is then obtained by the unitaries in the space of operators or, in more operational terms, as a unitary quantum circuit of the Liouvillian
\be
\Op(t)\equiv e^{i\mathcal{L}t}\Op(0)~.
\ee
In the next step, we would like to think about the Heisenberg operator as a state in a natural basis of orthonormal states $|\Op_n)$ associated with powers of the Liouvillian, the Krylov basis. The precise map between operators and these states will not be relevant for us, but interested readers can consult \eg  \cite{Magan:2020iac,Kar:2021nbm} for details using the GNS construction. 

There are a couple of relevant ingredients that we need to proceed. The first is the choice of the inner product.  It is commonly taken to have the Wightman form \cite{Parker:2018yvk}
\be \label{wightman}
(A|B)=\langle e^{ H\beta/2}A^\dagger e^{-H\beta/2}B \rangle_\beta~,
\ee
where, the right hand side is the thermal expectation value. With the inner product, we can then construct the Krylov basis following the Lanczos algorithm \cite{Lanczos:1950zz,Viswanath1990}. The key idea is to apply $\cL^n \Op$ recursively, followed by a Gram-Schmidt orthogonalization with all previously generated operators at each step.   We start by choosing the two orthonormal vectors 
\be
|\Op_0):=|\Op)~,\qquad |\Op_1):=b^{-1}_1\mathcal{L}|{\Op}_0)~,\qquad b_1^{-1}= (\Op_0 \cL | \cL \Op_0)~.
\ee
Then the Krylov basis is constructed iteratively by first computing 
\be
|A_n)=\mathcal{L}|\Op_{n-1})-b_{n-1}|\Op_{n-2})~,\label{AnDef}
\ee 
and then normalizing them
\be
|\Op_n)=b^{-1}_n|A_n)~,\qquad b_n=(A_n|A_n)^{1/2}~.
\ee
It is implicit that the operators/states are orthonormal, $(\O_m|\O_n)=\delta_{mn}$. 
The algorithm provides us not only with the set of Krylov vectors, $|\Op_n)$, but also with the so-called Lanczos coefficients, $b_n$. Upon rearranging terms in \eqref{AnDef}, we can see that the Liouvillian becomes a tridiagonal matrix in the Krylov basis
\begin{align}
	\cL_{mn} \equiv (\Op_m|\cL|\Op_n) = \begin{pmatrix}
		0 &b_1 &0 &0 &\cdots \\
		b_1 &0 &b_2 &0 &\cdots \\
		0 &b_2 &0 &b_3 &\cdots \\
		0 &0  &b_3 &0 &\ddots \\
		\vdots &\vdots  &\vdots &\ddots &\ddots \\
	\end{pmatrix}~. 
\end{align}
Finally, the Heisenberg operator/state \eqref{Heisenberg} in the Krylov basis becomes
\be\label{evo-opr}
|\Op(t))=\sum_{n}i^n\varphi_n(t)|\Op_n)~,
\ee
where, the coefficients, $\varphi_n$, satisfy a discrete Schr\"odinger equation
\be
\partial_t\varphi_n(t)=b_n\varphi_{n-1}(t)-b_{n+1}\varphi_{n+1}(t)\;.\label{SEq}
\ee
Once we have the Lanczos coefficients, solutions of this equation with boundary condition, $\varphi_n(t=0)=\delta_{n,0}$, fully determine the operator evolution. 

We can obtain a better physical intuition about the operator dynamics by considering a continuum limit of \eqref{SEq}, following  \cite{Parker:2018yvk,Barbon:2019wsy}. Let's define a lattice cut-off $\varepsilon$ and continuous variable $x=\varepsilon n$. The continuum version of the wavefunctions and the Lanczos coefficients then become  $\varphi(x,t)=\varphi_n(t)$ and $v(x)=2\varepsilon b(\varepsilon n)=2\varepsilon b_n$. These redefinitions render \eqref{SEq} into the following flow equation
\be
\partial_t\varphi(x,t)=\frac{1}{2\varepsilon}\left(v(x)\varphi(x-\varepsilon)-v(x+\varepsilon)\varphi(x+\varepsilon)\right).\label{SEqCont}
\ee
To the leading order in $\varepsilon$, this is  a first-order wave equation 
\begin{align}
	\label{flow-equation}
\partial_t\varphi(x,t)+v(x)\partial_x\varphi(x,t)+\frac{1}{2}v'(x)\varphi(x,t)=0~,
\end{align}
with position-dependent velocities $v(x)$ and mass $v'(x)/2$. Therefore, the Lanczos coefficients play the role of velocities for the spread of the initial operator wavefunction.  We will return to this interpretation in later sections.

The final step in this analysis is a definition of operator complexity called Krylov complexity. Intuitively, the operator evolution governed by the Schr\"odinger equation \eqref{SEq} can be thought of as a motion of a particle on a 1d lattice with sites labeled by different, $\varphi_n(t)$. As time progresses, the particle moves further along the lattice and the average position serves as a natural definition of operator's complexity. Concretely, the Krylov complexity (or K-complexity) is defined as
\begin{align}
	\label{K-com-def}
K_\Op (t)\equiv \sum_n n|\varphi_n(t)|^2~.
\end{align}
We also introduce the K-variance by which we can measure fluctuations around the average position in the 1d lattice. It is defined as follows 
\begin{align}\label{K-variance-def}
	\delta_{\cO}(t)^2 \equiv \frac{\sum_n n^2 |\varphi_n(t)|^2 - (\sum_n n |\varphi_n(t)|^2)^2}{(\sum_n n |\varphi_n(t)|^2)^2} =  \frac{\sum_n n^2 |\varphi_n(t)|^2 - K_\O(t)^2}{K_\O(t)^2}~. 
\end{align}
This quantity allows us to characterize the evolution further. 

To summarize, the above universal procedure allows us to construct the Krylov subspace for a given operator in a theory with Hamiltonian, $H$.  The Lanczos coefficients, $b_n$, and wavefunctions, $\varphi_n(t)$, linked by \eqref{SEq}, determine the operator evolution and the K-complexity \eqref{K-com-def}. This framework, together with various less universal q-complexities, can be used to classify and differentiate integrable and chaotic models \cite{Parker:2018yvk}. In particular, it is conjectured that Lanczos coefficients cannot have faster than linear growth in $n$. This upper bound should be saturated by chaotic systems; a canonical example is provided by the SYK model \cite{Sachdev:1992fk}. The linear growth of $b_n\sim a n$ translates to exponential growth of the K-complexity, with characteristic Lyapunov exponent $\lambda_{L}=2a$. This  operator growth hypothesis has been further studied and verified numerically in various examples \cite{Rabinovici:2020ryf,Barbon:2019wsy,Jian:2020qpp,Yin:2020oze,Magan:2020iac,Dymarsky:2019elm,Dymarsky:2021bjq,Carrega:2020jrk,Kim:2021okd,Kar:2021nbm,vonKeyserlingk:2017dyr,Nahum:2017yvy} .

Another recent progress in this topic that will play an important role in our work is \cite{Caputa:2021sib}. Motivated by parallel developments in CFT complexity \cite{Magan:2018nmu,Caputa:2018kdj,Chagnet:2021uvi,Koch:2021tvp}, this work focuses on Liouvillian circuits in systems governed by symmetries, such as CFTs. The main observation is that, the action of the Liouvillian in the Krylov basis 
\be \label{LActKS}
\mathcal{L}|\Op_n)=b_n|\Op_{n-1})+b_{n+1}|\Op_{n+1})~, 
\ee
allows for a representation of the Liouvillian in terms of the ladder operators of some Lie algebra governing the dynamics in Krylov subspace
\be\label{L-algo-0}
 \mathcal{L}=L_+ +L_-~,
\ee
such that
\be
L_-|\Op_n)=b_n|\Op_{n-1})~,\qquad L_+|\Op_n)=b_{n+1}|\Op_{n+1})~.
\ee
The advantage of this representation is that we can easily read off the Lanczos coefficients as well as Krylov basis  from representations of the  symmetry algebra. Moreover, a direct connection with coherent states allows us to extract the wavefunctions, $\varphi_n(t)$. In this light of symmetry, the operator growth can then be interepreted as geodesic motion in the information geometry (Fubini-Study metric), while the Krylov complexity is proportional to the volume in this geometry \cite{Caputa:2021sib}.

Let us review an example to illustrate the above ideas very briefly. The operator growth in the low-energy limit of the SYK model \cite{Sachdev:1992fk} belongs to a class governed by the  SL(2,R) symmetry algebra
\begin{align}\label{SL2RCom}
	[L_0,L_{\pm 1}]=\mp L_{\pm 1}~,\qquad [L_1,L_{-1}]=2L_0~,
\end{align}
for which the Liouvillian in the Krylov subspace can be written as, following \eqref{L-algo-0}, \cite{Caputa:2021sib}
\be\label{L-algo}
\mathcal{L}=\alpha\,(L_{-1}+L_{1})\;.
\ee
This choice is also closely related with the inner-product \eqref{wightman} and leads to the auto-correlator $\varphi_0(t)\sim \cosh^{-2h}(\alpha t)$ (see below).

The evolution protocol in the Krylov basis is then the following unitary action on the highest-weight state $\ket{h}=|\O_0)$
\be\label{L-algo-2}
|\O(t))=e^{i\alpha(L_{-1}+L_{1})t}\ket{h}~,
\ee
where $h$ is the eigenvalue of $L_0$ of $\ket{h}$. Such dynamics can in fact be interpreted as geodesic motion in the (hyperbolic) phase space spanned by SL(2,R) coherent states of Perelomov \cite{Perelomov:1971bd}.\footnote{The unitary circuit of the Liouvillian \eqref{L-algo-2} is a displacement operator of SL(2,R) (or SU(1,1)). See \cite{Haque:2021hyw} for recent study of the Euler-Arnold approach to circuits of this type.} This allows us to extract Lanczos coefficients and wavefunctions
\be\label{lanc-sl2}
b_n=\alpha\sqrt{n(2h+n-1)}~,\qquad \varphi_n(t)=\sqrt{\frac{\Gamma(2h+n)}{n!\Gamma(2h)}}\frac{\tanh^n(\alpha t)}{\cosh^{2h}(\alpha t)}~,
\ee
which agree with that found  using different methods in \cite{Parker:2018yvk} (their $\eta=2h$).
From these expressions, using \eqref{K-com-def}, we can compute Krylov complexity
\be
K_\Op=2h\sinh^2(\alpha t)\label{KrylovSL2R}~,
\ee
that is proportional to the volume in the hyperbolic disc from the origin up to radius $r=\alpha t$. Clearly, the Lanczos coefficients \eqref{lanc-sl2} grow linearly at large $n$, $b_n\sim \alpha n$, and K-complexity grows exponentially fast for late times with the characteristic Lyapunov exponent $\lambda_{L}=2\alpha$. In the above formulae, the coefficient $\alpha$ is not fixed by  symmetries and may depend on the properties of the Hamiltonian or the choice of the inner product. In the SYK example, we have $\alpha=\pi/\beta$; this leads to the same maximal Lyapunov exponent as from the OTOC, $\lambda_{L}=\lambda_{\rm OTOC}=2\pi/\beta$.

Last but not the least, we would like to mention that the starting point of the usual route to Krylov complexity in many-body physics is the auto-correlator
\be\label{AutoCor}
C(t)=\varphi_0(t)=(\Op(t)|\Op(0)).
\ee
From $C(t)$ and its Taylor expansion about $t=0$, one may extract Lanczos coefficients and solve the Schr\"odinger equation for $\varphi_n(t)$ \cite{Parker:2018yvk,Dymarsky:2021bjq}. We will argue that this seemingly universal shortcut overlooks some of the relevant and interesting details of the operator growth in systems with additional symmetries,   \eg systems with degeneracies like 2d CFTs,  while considering growth of primary or quasiprimary operators.

In what follows, we generalize this framework further in the context of 2d  CFTs. In particular, we will consider the Liouvillian \eqref{L-algo} (its representation in the Krylov space \eqref{LActKS}) as built from the global part of the Virasoro algebra and analyze the operator growth in this infinite dimensional Lie group.

\subsection{Oscillator formalism of the Virasoro algebra}
\label{subsec:osc}
In order to carry out the Lanczos algorithm, we need an orthogonal basis of operators. In 2d CFTs there are exact degeneracies arising from the descendant states. The descendants are excitations by the Virasoro generators on  primary states. These states are not orthonormal to each other by default and this poses an issue. A systematic solution to this is provided by the oscillator formalism of the Virasoro algebra and its representations.
In this subsection, we briefly review this formalism and how CFT states can be expressed in this framework. Further details of this formalism can be found in \cite[Appendix A]{Besken:2019bsu}. 

The Virasoro algebra is formed by the modes of the stress tensor
\be\label{Tz}
 T(z)=\sum_{n=-\infty}^\infty L_n z^{-n-2},
 \ee
 that obey the following commutation relation 
\begin{align}
	[L_m,L_n] = (m-n) L_{m+n} + \frac{c}{12}m(m^2 -1)~. 
\end{align}
The CFT Hilbert space is then organized into representations of two copies of the Virasoro algebra. The irreducible representations are primary states, $\ket{h}$, which are also related to the primary operators via the state-operator correspondence $\ket{h}=O(0)\ket{0}$. $h$ denotes the $L_0$ eigenvalue of the state or the holomorphic conformal dimension of the operator. The Verma module corresponding to each primary consists of descendant states that are created by the action of Virasoro generators on the primary, $L_{-1}^{m_1}L_{-2}^{m_2}\cdots \ket{h}$. 

As mentioned earlier, one disadvantage of working with the Virasoro basis of descendants, $L_{-1}^{m_1}L_{-2}^{m_2}\cdots \ket{h}$, is that the states in a given descendant level are not orthogonal (even in the standard sense, we are not referring to the inner-product \eqref{wightman} yet). This can be easily seen from the fact the off-diagonal elements of the Kac-matrix are generically non-vanishing. One can perform an orthogonalization procedure level-by-level but this gets incredibly laborious after a few low-lying descendant levels. An efficient means to proceed is offered by the oscillator basis that was initially developed to study 2d conformal blocks \cite{zamolodchikov1986two} (see also \cite{Besken:2019jyw}). In this basis, we represent a generic state $\ket{f}$ in the Verma module  by a Fock-Bargmann-like `wavefunction' $f(u)\equiv \vev{u|f}$, where $u$ denotes an infinite collection of oscillator variables $\{u_1,u_2,\cdots\}$.\footnote{The oscillator wavefunctions $\vev{u|f}$ are not to be confused with the wavefunctions $\varphi_n(t)$ of the Lanczos algorithm \eqref{evo-opr}. However, we will see there is a relation between these quantities as the story develops. } These $u_i$ are complex variables and $f(u)$ are holomorphic functions on $\mathbb{C}^\infty$. We denote the action of Virasoro generators as $l_nf(u) \equiv \vev{u|L_n|f}$. The operators $l_k$ have the following differential operator realizations
\begin{align}\label{vira}
	l_0 & = h+ \sum_{n=1}^\infty n u_n {\p\over\p u_n}~, \cr
	l_k &=\sum_{n=1}^\infty n u_n {\p\over\p u_{n+k}} -{1\over 4} \sum_{n=1}^{k-1} {\p^2 \over \p u_n \p u_{k-n} }+(\mu k+i\lambda){\p \over \p u_k}~,\quad\quad k>0 \\
	l_{-k} &= \sum_{n=1}^\infty (n+k) u_{n+k} {\p\over\p u_{n}} - \sum_{n=1}^{k-1}n(k-n) u_n u_{k-n}+2k(\mu k-i\lambda)u_k~,\quad\quad k>0~.\nn 
\end{align}
For $k=1$, the second term in the last two expressions are absent. 
A derivation of the above from quantization of the linear dilaton can be found in \cite{Besken:2019bsu}. 
In the above expressions the central charge, $c$, and the conformal dimension, $h$, are parametrized as
\begin{align}
	c=1+24\mu^2, \qquad h=\mu^2 +\lambda^2~. 
\end{align}
States in the conjugate representation $\bra{f}$ are denoted by $\overline{f(u)} =\vev{f|\bar{u}}$, where the bar acts on the oscillators as replacement by their anti-holomorphic counterparts $\bar{u}_n$.  The action of $L_n$ on $\bra{f}$ is as follows 
\begin{align}\label{conjugation}
	\vev{f|L_n|\bar{u}} = \overline{\vev{u|L_n|f}} = \overline{l_{-n}f(u)} = \bar{l}_{-n} \overline{f(u)}~. 
\end{align}
The bar also sends $i \mapsto -i$ in \eqref{vira} but leaves $\mu$ and $\lambda$ untouched. The oscillator wavefunctions are endowed with an inner product 
\begin{align}\label{InnerproductVir}
	\left( f(u), \,g(u)   \right) = \int [du] \overline{f(u)} g(u) ~, \qquad [du] = \prod_{n=1}^\infty d^2 u_n \frac{2n}{\pi} e^{-2n u_n \bar u_n}~. 
\end{align}
The specific measure meets the adjoint condition, $l_{-n}^\dagger =l_n$, thereby resulting in representations that are unitary.

The oscillator monomials,  $	u_1^{m_1} u_2^{m_2} \cdots $, with $\sum_j j m_j=N$,  then form an orthogonal basis of descendants at level $N$ of the primary $\ket{h}$. This is easily verified by the action of $l_0$, from equation \eqref{vira}, on the monomials. These monomials, however, are not normalized to unity. Their norms are as follows
\begin{align}\label{norm-desc}
	\left( u_1^{m_1} u_2^{m_2}  \cdots , u_1^{m_1} u_2^{m_2}  \cdots  \right) = S_{1,m_1} S_{2,m_2} \cdots ~,\qquad  S_{j,m} = \frac{m!}{(2j)^{m}}~. 
\end{align}
We can then define normalized descendants 
\begin{align}\label{Phi-def}
	\Phi_{\{m_i\}}(u)\equiv  \frac{u_1^{m_1} u_2^{m_2}  \cdots}{{\cN}_{\{m_i\}}}~, \qquad 	\cN_{\{m_j\}} = \sqrt{S_{1,m_1}S_{2,m_2}\cdots} = \left[\prod_{j=1}^{\infty} \frac{m_j!}{(2j)^{m_j}}\right]^{1/2}\!\!,
\end{align}
so that $(	\Phi_{\{m_i\}}(u), 	\Phi_{\{m_i\}}(u) )=1$ and  $(	\Phi_{\{m_i\}}(u), l_0 	\Phi_{\{m_i\}}(u) )=h+\sum_j jm_j=h+N$. It is clear from these relations that the orthogonal descendants are labeled by integer partitions of the descendant level $N$.

The orthonormal basis of descendant states \eqref{Phi-def} will extensively feature in our computations below. Note that since we shall be working with this orthonormal basis per se, the Gram-Schmidt step of the Lanczos algorithm will be redundant. On a related note, since there are exact  degeneracies, it is impossible to `compress' information of the $N$'th step of the Liouvillian evolution into a single operator/state. In contrast to the standard Lanczos algorithm, this is a key difference. 

\subsection{Descendants, Young diagrams and the Young's lattice}
\label{subsec:young}

We have just noticed that the orthogonal descendants \eqref{Phi-def} have a one-to-one correspondence to integer partitions of the descendant level. The descendants can be labeled explicitly by the integer partition and can be written as follows
\begin{align}\label{desc-notation}
	 \ket{1^{m_1}2^{m_2}\cdots} 
	 \mapsto  \Phi_{\{m_i\}}(u)
	 ~, \qquad \sum_{j=1}^{N} jm_j =N~. 
\end{align}
Young diagrams are a useful way to visualize integer partitions. The total number of boxes in the Young diagram simply corresponds to the descendant level $N$ and the specific arrangement of boxes give the specific partitions or orthonormal descendants at a given level. For instance, at level 3, we have the following states and their corresponding Young diagrams
\begin{align}
	\ket{1^3} \mapsto  \tableau{1 1 1}, \quad \ket{1^1 2^1} \mapsto  \tableau{2 1}, \quad \ket{3^1} \mapsto  \tableau{3}~. 
\end{align}
In terms of the oscillator variables, the above states are proportional to $u^3_1$, $u_1u_2$ and $u_3$ respectively.

\begin{figure}[t!]
	\centering
	\includegraphics[width=1\linewidth]{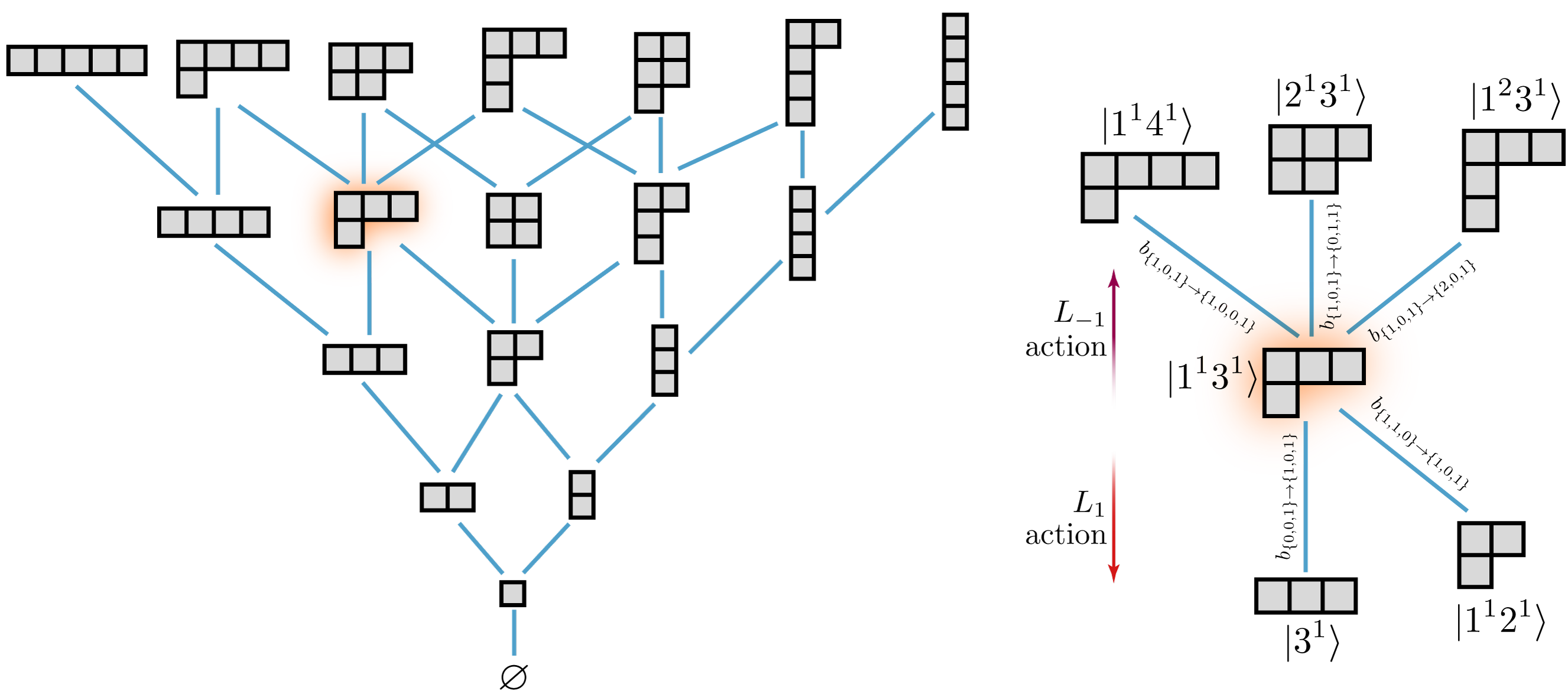}
	\caption{[Left] The Young's lattice denoting descendant states till level 5. [Right] Focusing on a specific vertex/descendant and its edges weighted by the Lanczos coefficients.}
	\label{fig:younglattice-main}
\end{figure}

\begin{figure}[t!]
	\centering
	\includegraphics[width=0.8\linewidth]{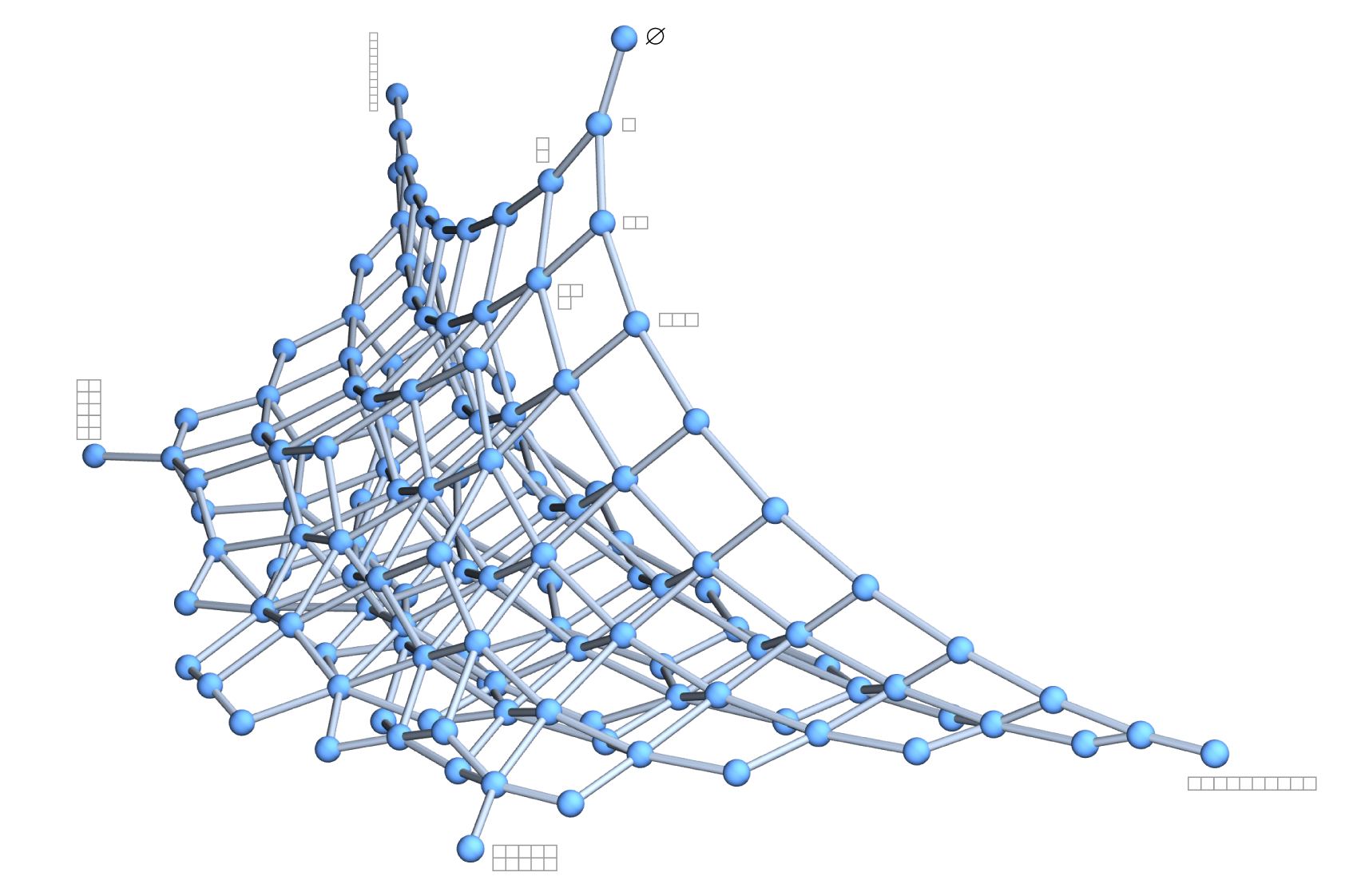}
	\caption{A 3d view of the Young's lattice till level 10. Some of the vertices have been marked by the Young diagrams. (This can be generated in Mathematica using the function \texttt{GraphPlot3D} with the adjacency matrix of the Young's lattice as the input.)}
	\label{fig:3dyoung}
\end{figure}

We now introduce the Young's lattice. This is a graph (in the mathematical sense of the term) which have Young diagrams at its vertices, see \cref{fig:younglattice-main,fig:3dyoung}. Each layer consists of Young diagrams corresponding to partitions of a fixed integer. The edges of the graph connect two Young diagrams which can be related by the addition or removal of a single box. This construction will turn out to be very useful to describe the Lanczos sequence of primary operators.\footnote{Strictly speaking, we should be considering the Young's lattice as a directed graph. This is because we start with the primary state at $t=0$ and the evolution proceeds into higher level descendants. } The layers corresponds to the descendant levels. The Liouvillian, defined in \eqref{L-algo}, precisely performs the task of adding/removing a single box. This can be manifestly seen from the differential operator realizations of $l_{\pm1}$ from \cref{vira}
\begin{align}\label{lpm1}
	l_1 &=\sum_{n=1}^\infty n u_n {\p\over\p u_{n+1}} +(\mu +i\lambda){\p \over \p u_1}~, \qquad 
	l_{-1} = \sum_{n=1}^\infty (n+1) u_{n+1} {\p\over\p u_{n}}  +2(\mu -i\lambda)u_1~.
\end{align}
Let's consider the action of these operators on an arbitrary descendant state \eqref{Phi-def}. 
The action of $l_{1}$ on the monomial $u_1^{m_1}u_2^{m_2}\cdots$ is equivalent to removal of a box from one of the inner corners of the Young diagram, while the action of  $l_{-1}$ corresponds to addition of a box to one of the outer corners. For example, consider the descendant at level 8
\begin{align} \label{l8eg}
\ket{1^1 2^2 3^1}	\mapsto \tableau{3 2 2 1}
\end{align} 
The action of $l_1$ leads to a superposition of the following three states at level 7
\begin{align}
\ket{1^1 2^3} \mapsto	\tableau{2 2 2 1}~, \quad \ket{1^2 2^1 3^1} \mapsto  \tableau{3 2 1 1}~, \quad \ket{2^2 3^1} \mapsto \tableau{3 2 2} ~. 
\end{align}
On the other hand, the action of $l_{-1}$ creates a superposition of four states at level 9
\begin{align}\label{l9eg}
	\ket{1^1 2^2 4^1}	\mapsto \tableau{4 2 2 1}~, \quad  \ket{1^12^1 3^2}	\mapsto \tableau{3 3 2 1}~, \quad \ket{ 2^3 3^1}	\mapsto \tableau{3 2 2 2}~, \quad \ket{1^2 2^2 3^1}	\mapsto \tableau{3 2 2 1 1}~. 
\end{align}
Therefore, the state \eqref{l8eg} has three edges from the layer below and four edges to the layer above; see \cref{fig:younglattice-main} (right) for another example.
In general, the number of ways of adding a single box is always one greater than the number of ways of removing a box. In graph theory,  this is often phrased as: each vertex of the Young's lattice has degree $2M+1$, and it consists of $M$  {predecessors} and $M+1$  {successors}. 

Let us now point out a couple of combinatorical properties about the Young's lattice that will play a role in the next section. First, the number of Young diagrams at the $N$'th layer is simply the number of integer partitions $p(N)$. In CFT terms, this is the number of descendants at each level and this can be seen from the Virasoro character of the primary 
\begin{align}\label{char}
	\chi_h(q) = \frac{q^{h-\frac{c-1}{24}}}{\eta(q)} = q^{h-\frac{c}{24}} \prod_{n=1}^{\infty} \frac{1}{1-q^n}~. 
\end{align}
It is crucial that we are working with irrational CFTs with no null-states in non-vacuum modules, so that the above form of the character can be used. 
The infinite product in \eqref{char} serves as a generating function for the number of partitions $p(N)$. Second, we can also calculate the number of edges, $B(N+1)$, that connect the $N$'th layer to the $(N+1)$'th layer of the   lattice. In order to find this, we use the fact that the number of successors of each vertex is one more than the number of predecessors. This leads to a recurrence relation between the number of edges. Let's write it out in words first:
\begin{align}
	&\text{ $\#$ of edges to level $N+1$ = $\#$ of edges to level $N$} \\
	&\qquad\qquad\qquad\qquad\quad\qquad\qquad\text{~+ one additional edge from each vertex at level $N$ }.\nn 
\end{align}
Symbolically, this is the recursion relation
\begin{align}
	B(N+1) = B(N)+p(N)	~,
\end{align}
where, $p(N)$, is the number of integer partitions of $N$ counting the additional edges from each Young diagram. The seed condition for this recursion is $B(0)=1$, as there is one edge from $\varnothing$ to $\tableau{1}$. The solution to the recurrence relation then yields the cumulative sum of integer partitions\footnote{See also \cite[Exercise 1.71 \& Theorem 3.21.11]{stanley2011enumerative}.}
\begin{align}\label{branches}
	B(N)=	\sum_{j=0}^{N} p(j) ~, 
\end{align}
with $p(0)=1$.
A generating function counting the edges between the layers is then given by
\begin{align}
	\sum_{n=0}^\infty B(N)q^n = \frac{1}{1-q}\prod_{j=1}^\infty \frac{1}{1-q^j}~. 
\end{align}
The Lanczos coefficients are weights of the edges of the Young's lattice (see \cref{fig:younglattice-main}) and the $B(N)$'s are precisely the number of non-vanishing Lanczos coefficients while transitioning between two adjacent descendant levels. Finally, the late time regime of the Lanczos algorithm maps to the regime of high-level descendants. These corresponds to Young diagrams with large number of boxes that lie in the asymptotic regime of the Young's lattice. We explore these aspects further in the next section.

\section{Operator growth as spreading in the Young's lattice}
\label{sec:liov}
With the general formalism for operator growth and tools for 2d CFT in place, we now study the evolution of a primary operator $\O$ (having holomorphic conformal dimension $h$) under the Liouvillian evolution \eqref{L-algo}. 

\subsection{Lanczos algorithm and the Young's lattice}
\label{subsec:lanczos}
The Liouvillian evolution of the primary state will create a linear combination of descendant states within the same Verma module. The coefficients of this linear combination will be time-dependent and will be labelled by integer partitions of the descendant level $N$. We will see the details of this linear superposition in the next section. For now, let us focus on how the Lanczos algorithm works. We shall focus on the holomorphic sector of CFT; the generalization to the anti-holomorphic sector is simply a copy of what happens for the holomorphic one.

\begin{figure}[!t]
	\centering
	\includegraphics[width=\textwidth]{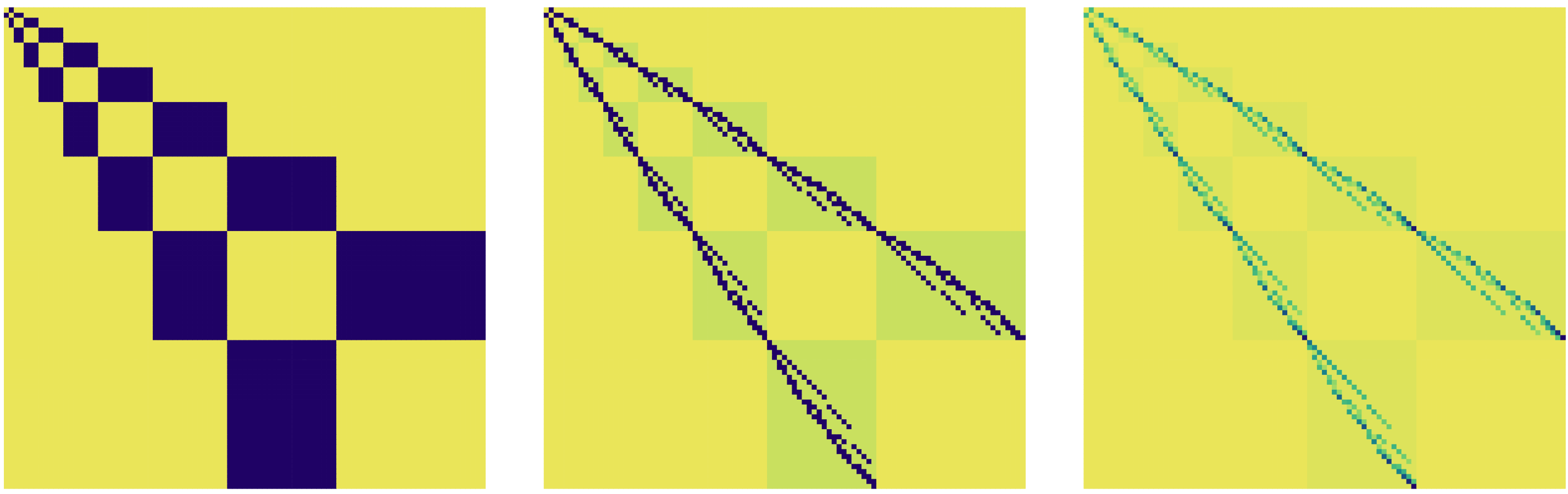}
	\caption{[Left] The naive tri-diagonal form of the Liouvillian in the orthonormal basis of descendants. Each block shows a Lanczos matrix of dimensions $p(N)\times p(N\pm1)$.  [Centre] The adjacency matrix of the Young's lattice, the black elements indicate that two vertices are nearest neighbours and will have a non-zero Lanczos elements. [Right] Matrix plot of magnitudes of the Lanczos elements for $h=10$ and $c=5$.\protect\footnotemark }
	\label{fig:matrix-trio}
\end{figure}
\begin{figure}[!t]
	\centering
	\includegraphics[width=\linewidth]{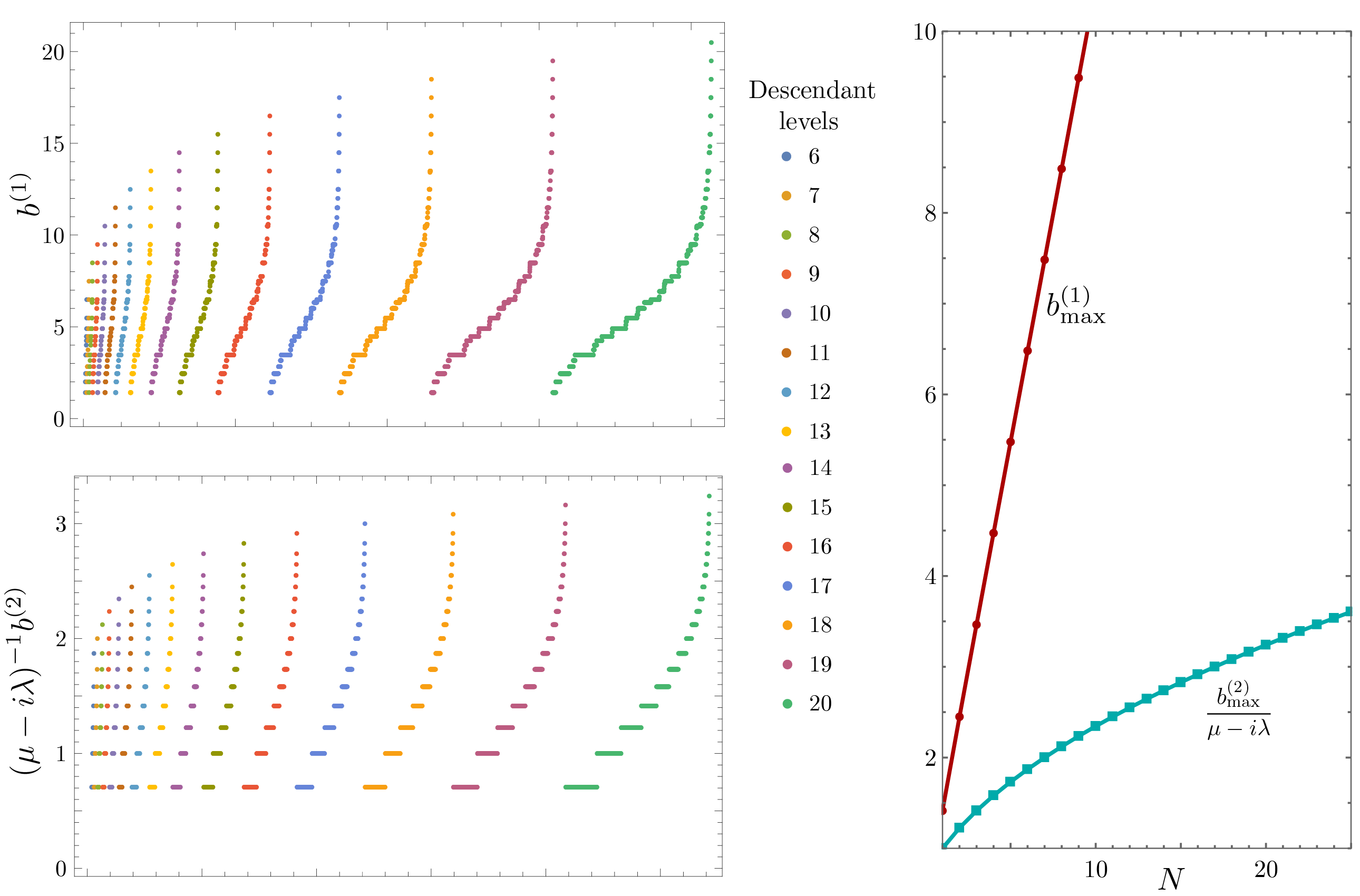}
	\caption{[Left] Plots of type-1 and type-2 Lanczos elements \cref{type1,type2} from descendant levels 6 to 20. We have set $\alpha=1$ and the data for each descendant level is sorted in the ascending order. [Right] Maximum values of the Lanczos element from each descendant level while the maximal type-1 elements have an asymptotic linear growth, while the maximal type-2 elements have $\sqrt{N}$ growth. These are  derived analytically in \cref{maximal-Lanc,col-Lanczos}. }
	\label{fig:b-scatter}
\end{figure}

\footnotetext{The faded rectangles in the second and third figures are meant to serve as a visual guide, and do not indicate non-zero values. In the third figure, rescaled and offsetted values have been used for the sake of clarity; please see \cref{fig:b-scatter} for actual numerical values.  }
The Lanczos algorithm, \eqref{L-algo} and \eqref{L-algo-2}, for the Virasoro module proceeds in a slightly generalized manner as we have exact degeneracies. At the $N$'th step of the process the Liouvillian, $\cL \equiv \alpha (L_{-1}+L_1)$, acts on descendants at level $N$ and creates a superpositions of descendants at levels $N-1$ and $N+1$. In the orthonormal oscillator basis \eqref{Phi-def}, we can write the action on an arbitrary descendant as follows 
\begin{align}\label{Liov-Vir}
\bra{u}	\cL  \Phi_{\{m_k\}} \rangle =\alpha   (l_{-1}+l_1)\Phi_{\{m_k\}} (u) =\!\!\!\!\sum_{\sum jr_j = N+1 }\!\!\! b_{\{m_k\}\to\{r_j\}} \Phi_{\{r_j\}} (u) + \!\!\!\!\sum_{\sum js_j = N-1 }\!\!\! b_{ {\{s_j\}}\to \{m_k\}} \Phi_{\{s_j\}}(u) ~,
\end{align}
for $\sum km_k=N$  -- see \cref{fig:younglattice-main} (right) for a schematic illustration of the above equation. 
The first term on RHS arises from the $l_{-1}$ action, whilst the second is from the action of $l_1$.
Here, we have denoted the partitions of $N+1$ by $\{r_j\}$  and those of $N-1$ by $\{s_j\}$. As remarked earlier, there is no way to compress the information of the $N$'th step of the evolution into a single state/operator due to exact degeneracies in a Verma module. Therefore, the Lanczos coefficients of the non-degenerate case  generalize to matrices, $b_{\{m_k\}\to\{r_j\}}  $. We will refer to this as the Lanczos matrix. It is a rectangular matrix with dimensions $p(N) \times p(N+1)$ as the elements of this matrix are labelled by integer partitions of $N$ and $N+1$. Equivalently we can label the matrix elements by Young diagrams of $N$ vs.~$N+1$ boxes. 
The action of $l_{\pm 1}$ connects adjacent descendant levels gives rise to a block tridiagonal structure of the full Liouvillian in the oscillator basis; each block corresponds to the rectangular Lanczos matrix between two adjacent descendant levels -- see \cref{fig:matrix-trio} (left). 
We will see in a moment that each block is sparse, with the non-zero entries lying somewhat close to the diagonal.

Our task now is to find the values of the elements of the Lanczos matrix, $b_{\{m_k\}\to\{r_j\}}$. It will suffice to consider the action of $l_{-1}$ on an arbitrary descendant state to get this information. From \eqref{Liov-Vir}, elements of the Lanczos matrix can be thought of as transition amplitudes of the following kind
\begin{align}\label{transit}
	b_{\{m_k\}\to\{r_j\}} = \left(\Phi_{\{m_k\}}(u) , \alpha \, l_{-1} \Phi_{\{r_j\}} (u)    \right)~. 
\end{align}
This is the amplitude for a descendant state of level $N$ to go to one in  level $N+1$. Moreover, this corresponds to adding a single box to a specific Young diagram, \eg \eqref{l8eg} to \eqref{l9eg}, and \cref{fig:younglattice-main} (right). 
The amplitude for the reverse process is given by the $l_1$ action and this corresponds to deleting a box from the Young diagram; such amplitudes can be found simply by conjugation \cref{conjugation}, as $l_1= (l_{-1})^\dagger$ for unitary representations.

Let us now find the Lanczos matrices. We use the differential operator realization of the generator $l_{-1}$ from \eqref{vira} and act it on a normalized descendant state of level $N$ \eqref{Phi-def} 
\begin{align}\label{lm-action}
	l_{-1}\Phi_{\{m_k\}}(u)  
	= & \sum_{n=1}^N (n+1)m_n   {\cN_{\{m_1,m_2,\cdots,m_n-1,m_{n+1}+1,\cdots\}}\over   \cN_{\{m_k\}}   }(u)   \Phi_{\{m_1,m_2,\cdots,m_n-1,m_{n+1}+1,\cdots\}}(u)  \nn \\
	&+2(\mu -i\lambda)\frac{\cN_{\{m_1+1,m_2,\cdots \}}}{\cN_{\{m_k\}}}    \Phi_{\{m_1+1,m_2,\cdots \}}(u)  ~. 
\end{align}
The ratios of the normalization coefficients can be simplified further using \eqref{Phi-def} and we have the result
\begin{align}\label{lm-action-2}
	l_{-1}\Phi_{\{m_k\}}(u)
	= & \sum_{n=1}^N \sqrt{n(n+1)m_n (m_{n+1}+1)} \Phi_{\{m_1,m_2,\cdots,m_n-1,m_{n+1}+1,\cdots\}}(u)  
	\nn \\&
	+ (\mu-i\lambda) \sqrt{2(m_1+1)}   \Phi_{\{m_1+1,m_2,\cdots \}}(u)  ~. 
\end{align}
As the above equation is written fully using orthonormal states, we can directly read-off the elements of the Lanczos matrix \eqref{transit}. Since there are two types of terms above, those with a $(h,c)$ dependence and those without, we call these as type-1 and type-2 Lanczos elements. These are 
\begin{align}
	&	\text{Type 1:} \qquad b^{(1)}_{\{m_j\}\to \{m_1,m_2,\cdots,m_n-1,m_{n+1}+1,\cdots\}} ~=~ \alpha\sqrt{n(n+1)m_n (m_{n+1}+1)} ~,   \label{type1} \\
	& \text{Type 2:} \qquad b^{(2)}_{\{m_j\}\to \{m_1+1,m_2,\cdots \}} ~=~ \alpha (\mu-i\lambda) \sqrt{2(m_1+1)}  ~. \label{type2}
\end{align}
A plot of these coefficients for some descendant levels is shown in \cref{fig:b-scatter} (left) and a matrix plot of the magnitudes for $h=10$ and $c=5$ is shown in \cref{fig:matrix-trio} (right).

The matrix plot  in \cref{fig:matrix-trio} (right) reveals that a large fraction of the elements in the Lanczos matrices,  $b_{\{m_k\}\to\{r_j\}}$, are zero. 
We can count the number of non-zero elements of the Lanczos matrix between two consecutive descendant levels. Recall that the action of $l_{-1}$ is equivalent to adding a single box to a Young diagram and the Lanczos elements are weights of the edges of the Young's lattice -- \cref{fig:younglattice-main} (right). Therefore, the number of non-vanishing elements of the Lanczos matrix, \eqref{type1} and \eqref{type2}, is precisely given by the number of edges between two layers of the Young's lattice \eqref{branches} -- also see \cref{fig:younglattice-main}. In other words, we get non-zero entries whenever two Young's diagrams are nearest neighbours in the lattice. In graph theory parlance, the adjacency matrix distinguishes which vertices are nearest neighbours (or adjacent) and those which are not by 1's and 0's respectively. The adjacency matrix for the Young's lattice is shown in \cref{fig:matrix-trio} (centre). 

For large descendant levels, the non-zero entries actually constitute a very small fraction of the Lanczos matrix. This can be seen as follows. In the asymptotic regime, the number of edges/links  between two adjacent layers of the Young's lattice  is, from \eqref{branches}, 
\begin{align}
	B(N\to \infty) \approx \int_0^N dn\,p(n) \approx \int_0^N dn  \,\frac{e^{\pi \sqrt{2n/3}}}{4n\sqrt{3}} \approx \frac{ e^{\pi \sqrt{2N/3}}  }{ 2\pi \sqrt{2N}   }~. 
\end{align}
where, we used the Hardy-Ramanujan formula for $p(n)$. On the other hand the dimension of the Lanczos matrix for large $N$ is essentially the square of the Hardy-Ramanujan growth
\begin{align}
\dim[b_{\{m_k\}\to \{r_j\}}]	= p(N) \times p(N+1) \stackrel{N\to \infty}{\approx}  \frac{e^{2\pi \sqrt{2N/3}}}{48N^2}~. 
\end{align}
Hence, the fraction of non-zero elements is exponentially suppressed $\sim e^{-\pi\sqrt{2N/3} }$. We now study these non-vanishing  elements,  \eqref{type1} and \eqref{type2}, and find their growth properties.

\subsubsection*{Maximal growth -- single row diagrams}
It can be seen that specific type-1 elements \eqref{type1} saturate the conjectured bound  on maximal growth \cite{Parker:2018yvk}. Consider level-$N$ descendant states of the kind $\ket{N^1}$ in the notation of \eqref{desc-notation}. In the oscillator basis, these are states proportional to $u_N$. In other words, these are Young diagrams with a single row and lie at the rightmost edge of the Young's lattice, \cref{fig:younglattice-main}; \ie this is the path
\begin{align}\label{col-path}
\varnothing \to	\tableau{1}  \to	\tableau{2} \to	\tableau{3} \to	\tableau{4} \to	\tableau{5} \to \cdots~.  
\end{align}
 Therefore, the integer partition, $\sum jm_j=N$, has $m_N=1$ and all other $m_{j\neq N}=0$. The $l_{-1}$ action on this state creates a similar descendant, $\ket{(N+1)^1}$, of the level $N+1$ which has $m_{N+1}=1$ and $m_{j\neq N+1}=0$, in addition to some other states. Let's consider type-1 element  \eqref{type1} corresponding to these two similar descendants of adjacent levels (\eg $\ket{4^1}\to \ket{5^1}$ or  $\tableau{4} \to \tableau{5}$)
\begin{align}\label{maximal-Lanc}
	 b^{(1)}_{\{0,\cdots,0,1_N,0,\cdots \}\to \{0,\cdots,0,1_{N+1},0,\cdots \} }= \alpha\sqrt{N(N+1)}~ \stackrel{N\to \infty}{\approx}~ \alpha N ~. 
 \end{align}
The notation $1_N$ indicates `1' at the $N$'th position in the set. Therefore, we find asymptotically linear growth along the rightmost edge of the Young's lattice -- this is also verified numerically in \cref{fig:b-scatter} (right, red curve). Hence, transitions between descendants of the kind \eqref{col-path} saturate the upper bound on Lanczos coefficients conjectured  in \cite{Parker:2018yvk}. This is the one of the key results of our work.  As the Lanczos coefficients have an interpretation as velocities for operator growth \eqref{flow-equation}, we can infer that the channel \eqref{col-path} has the fastest spread of information about the primary operator. This linear growth of Lanczos coefficients manifests itself in the exponential growth of the total Krylov complexity and we will see that in the next section.

\subsubsection*{Growth for single column diagrams}

We now consider the type-2 elements \eqref{type2} and specifically the transitions between descendants of the kind $\ket{1^N} \to \ket{1^{N+1}}$; this corresponds to the integer partitions $m_1=N,m_{j\neq1}=0$ and $m_1=N+1,m_{j\neq1}=0$ respectively. The oscillator realization these states is proportional to $u_1^N$ and the Young diagrams corresponding to these have a single column
\begin{align}
	\varnothing \to \tableau{1} \to \tableau{1 1} \to \tableau{1 1 1}  \to \tableau{1 1 1 1} \to \tableau{1 1 1 1 1} \to \cdots ~. 
\end{align}
The Lanczos elements corresponding to above transitions are, from \eqref{type2}
\begin{align}\label{col-Lanczos}
	b^{(2)}_{\{N,0,0,\cdots \}\to \{ {N+1},0,0,\cdots \} }= \alpha (\mu -i \lambda)\sqrt{2(N+1)}~ \stackrel{N\to \infty}{\approx}~\alpha (\mu -i \lambda)\sqrt{2N}~. 
\end{align}
Therefore, along the leftmost edge of the Young's lattice we find $N^{1/2}$ growth. In fact, this is fastest possible growth for the type-2 coefficients \eqref{type2} -- see \cref{fig:b-scatter} (right, green curve). Note that, if the conformal dimension of the primary is large, for intermediate descendant levels $(N\sim h)$ there is a competition between the above elements \eqref{col-Lanczos} and those from \eqref{maximal-Lanc}. Comparing the modulus of the elements, we see that the elements of \eqref{maximal-Lanc} ultimately win in the  regime 
$N\gg 2h$. 

\subsubsection*{Lanczos elements for hook diagrams}

A case that interpolates between single row/column diagrams are  hook diagrams. These are descendants of the form $\ket{1^{N-r}r^1}$ and, hence, have $m_r=1$ (with $r=0$ or $2\leq r\leq N$) and $m_1=N-rm_r=N-r$ and all other $m_i=0$; $r$ is the position of the hook when counting boxes from the right. Here's an example, from descendant level 9
\begin{align}
\ket{1^4 5^1} \mapsto	\tableau{5 1 1 1 1}~. 
\end{align}
The $l_{1}$ action is given by (the notation $1_r$ denotes `1' at the $r$th position)
\begin{align}\label{hook}
	l_{-1}\Phi_{\{N-r,0,\cdots,0,1_r,0,\cdots\}}
	= &  \sqrt{r(r+1)} \, \Phi_{\{N-r,0,\cdots,0,1_{r+1},0\cdots\}}
	+ \sqrt{2(N-r)} \Phi_{\{N-r-1,1,0,\cdots,0,1_r,0\cdots \}}
	\nn \\ &
	+(\mu-i\lambda) \sqrt{2(N-r+1)}  \,  \Phi_{\{N-r+1,0,\cdots,0,1_r,0,\cdots \}}~. 
\end{align}
As expected, the result interpolates between the two extreme cases above: for $r=N$ we get \eqref{maximal-Lanc}  from the 1st term, while for $r=0$ we get \eqref{col-Lanczos}  from the last term.  Furthermore, the terms on the RHS of \eqref{hook} is linear combination of the descendants of the following types
\begin{align}
	\ket{1^4 6^1}\to \tableau{6 1 1 1 1}, \quad \ket{1^3 2^1 5^1}\to	\tableau{5 2 1 1 1}, \quad  ~ \ket{1^5 5^1}\to\tableau{5 1 1 1 1 1}~. 
\end{align}
The first diagram has one box added in the first row, the second has one box added in the second column and the third has one box added to the first column. 

\subsubsection*{Growth for typical descendants}

The cases considered above are highly atypical in the regime of high descendant levels. The notion of typicality needs to be defined with respect to some observable. We consider stress tensor correlators, for which thermal expectation values are reproduced by typical  high-level descendants of a heavy primary \cite{Datta:2019jeo}. The numbers $m_j$
of the integer partition $\sum_j j m_j=N$ that are distributed according to the Boltzmann distribution and have a Bose-Einstein mean 
\begin{align}
	\overline{	m_j } = \frac{q^j}{1-q^j}, \quad \text{with } q= e^{-\pi / \sqrt{6N}}. 
\end{align}
The Lanczos coefficients of this typical state are, from \eqref{type1} and \eqref{type2} 
\begin{align}
	&	\text{Type 1:} \qquad b^{(1)}_{\{\overline{m_j}\}\to \{ \overline{m_1}, \overline{m_2},\cdots, \overline{m_n}-1,\overline{m_{n+1}}+1,\cdots\}}  \approx \left[ n(n+1) \frac{q^{2n+1}}{(1-q^n)(1-q^{n+1})} \right]^{1/2},\\
	& \text{Type 2:} \qquad b^{(2)}_{\{\overline{m_j}\}\to \{\overline{m_1}+1,\overline{m_2},\cdots \}}   \approx (\mu-i\lambda) \left[\frac{2q}{1-q}\right]^{1/2}.
\end{align}
Now, let's focus on transition amplitudes with $n\ll N$, \ie we are looking at specific transitions of this type or specific terms of this kind in \eqref{lm-action-2}. We then make further approximations and we get the following estimates
\begin{align}
	b^{(1)}_{\rm typ}   &\approx \frac{\sqrt{6N}}{\pi}~,
	\qquad 
	b^{(2)}_{\rm typ}  \approx 
	  (\mu-i\lambda) \frac{(6N)^{1/4}}{\sqrt{\pi}}~. 
\end{align}
Therefore, type-1 and 2 Lanczos coefficients for typical states scale as $N^{1/2}$ and $N^{1/4}$ respectively.\footnote{Note that one has to be careful with the interpretation here: these are the  {Lanczos coefficients \textit{for} typical states} (in addition to the small $n$ specification) and  {not}  {the typical value for the Lanczos coefficient}. Typical values for these coefficients do not make sense in present context as the variance/spread is high.} Interestingly, this square-root slow-down of Lanczos coefficients along the evolution is expected on various grounds (see \eg \cite{Barbon:2019wsy}) and corresponds to the period of sub-exponential or power-law growth of Krylov complexity. The $\sqrt{N}$ growth is seen very explicitly in our example and we can pin-down the typical operators responsible for this mechanism. This is also one of our main findings.

\subsection{Asymptotics of operator spreading}
\label{subsec:asymp}

We have seen that spreading of the primary operator into the bath of descendants can be very concretely represented using  the Young's lattice. The regime of late times, gets mapped to high descendant levels or layers of large heights in the Young's lattice. This brings us to a natural question: what are the number of paths (or histories) to a high-level descendant? This quantity should be completely determined by combinatorics and can potentially furnish an estimate of which states are probable and which aren't at late times. Furthermore, the number of histories also crucially enters as an input in studies of operator growth for the bounding norms of operators and their commutators, see \eg \cite{Avdoshkin:2019trj,Chen:2019klo,Cao:2020zls}.

From the perspective of the Young's diagrams, the answer to this question is exactly the number of allowed fillings of the Young diagram or, in more technical terms, the number of possible {standard Young tableaux of a specified shape}. The entries within the boxes of a standard Young tableau actually correspond to the sequence in which single boxes can be added to $\varnothing$ to arrive at that specific shape of the Young diagram. Therefore, the number of allowed Young tableaux is the same as the number of paths starting from $\varnothing$ to a specific   diagram on the Young's lattice.  This is counted by the Young-Frobenius formula or equivalently by hook's formula\footnote{This also counts the dimensions of irreps of the symmetric group, $S_N$. }
\begin{align}\label{hook-formula}
	f^m = N! \det \left[1\over (m_j -j +k)!\right] = \frac{N!}{\prod h_{m}(j,k)}~, \qquad m \equiv \{m_j\}\,,\,  \sum_j jm_j = N~. 
\end{align}
The determinant in \eqref{hook-formula} is of a $l\times l$ matrix, where $l$ is the length of the partition, and it also uses the convention $1/r!=0$ if $r<0$. 

Now let's consider the late time regime or, equivalently, the asymptotics of the number of ways to reach a high-level descendant or to a Young diagram with a large number of boxes. For large $N$, it was proved by Vershik and Kerov \cite{vershik1985asymptotic} (see also \cite{logan1977variational}) that the largest $f^m$ for a fixed $N$ is bounded from both sides, as follows
\begin{align}
	\sqrt{N!}\, e^{-c_1\sqrt{N}[1+O(1)] }\leq \max[f^m] \leq \sqrt{N!}\, e^{-c_2\sqrt{N}[1+O(1)]} ~. 
\end{align}
where, $c_1 = \pi/\sqrt{6} = 1.2825..,~c_2 = \pi^{-1} - 2\pi^{-2} = 0.1157..$\,. 
This result can be directly utilized to provide an upper bound on the number of paths leading to specific Young diagram in the Young's lattice. Let's consider the upper bound 
\begin{align}\label{fk-bound}
	f^m \leq \sqrt{N!}\,  e^{-c_2\sqrt{N}[1+O(1)]}~, \qquad \implies~  \log f^m \leq \frac{1}{2} \log N!  - c_2 \sqrt{N}[1+O(1)]~. 
\end{align}
We now situate this into the context of operator growth. An upper bound on the average number of paths or histories, after Liouvillian evolution by $N$ steps,  has been conjectured to be (see \cite[eq.\,(29)]{Avdoshkin:2019trj})
\begin{align}
	h(N) \leq C' \frac{N!}{\epsilon^N}~, \qquad \implies~\log h(N) \leq   \log N!  - N \log \epsilon + \log C' ~. 
\end{align}
where $\epsilon >1$ and $C'>1$. The upper bound we get for the Young's lattice \eqref{fk-bound} is stronger than the above. The result \eqref{fk-bound} can be potentially applied, along with some other non-universal inputs, to bound norms of operators and square of commutators in 2d CFTs. We do not pursue this direction further here, and now turn to the details of the state that resulted from the Liouvillian evolution.

\section{Characterizing the evolved state}
\label{sec:evo}

In this section we study the properties of the evolved primary state under the protocol \eqref{L-algo}. We shall find an exact analytic expression for the evolved state. This will allow us to characterize the state further through the K-complexity and Renyi entropies. 

\subsection{The evolved state in closed form}
\label{subsec:evo}
We start by deriving a closed form expression for the evolved state in the oscillator basis. First, the time evolution can be simplified to some extent by using the Baker–Campbell–Hausdorff relation for the SL(2,R) generators. More generally, we write
\begin{align}\label{evo-action}
	e^{(\xi L_{-1} -\bar \xi L_{1})} \O(0) \ket{0}&= e^{\alpha_{-}  L_{-1}}e^{\alpha_0  L_0}e^{\alpha_+  L_1} \O(0) \ket{0} 
	= e^{\alpha_0h} e^{\alpha_{-} L_{-1}} \O(0) \ket{0}~,
\end{align}
where we have used the relations $L_{1}\O(0)\ket{0}=0$, $L_0\O(0)\ket{0}=h\O(0)\ket{0}$ and the following definitions
\begin{align}\label{alpha-defs}
	\xi = re^{i\phi}\qquad \alpha_{\pm} = \mp e^{\mp i\phi} \tanh (r)~, \quad \alpha_0 = -2 \log \cosh (r)~. 
\end{align}
The factorized form with the SL(2,R) generators in \eqref{evo-action} implies that the time evolution protocol is a combination of special conformal transformations (which act trivially), dilatations and translations. More precisely, the evolution of the operator in the Krylov basis corresponds to a certain trajectory in these states (\ie phase space) parametrized by \cite{Caputa:2021sib}
\be\label{TimeDepPar}
\xi=i\alpha t,\quad\text{or}\quad r=\alpha t,~ \phi=\pi/2~.
\ee
Using the fact that $L_{-1}$ is the generator of translations we can write \eqref{evo-action} as
\begin{align}\label{evo2}
	\ket{\O(t)}\equiv e^{i\alpha(L_{-1} + L_{1})t} \O(0) \ket{0}&= e^{\alpha_0 h} \O(\alpha_-)\ket{0}~,\qquad \alpha_-=i\tanh(\alpha t)~,
\end{align}
where $\Op(\alpha_-)$ is a local primary operator inserted in $\alpha_-$.

It is now time to deploy the oscillator basis. The object $\O(z)\ket{0}$ is known in closed form \cite[eq.\,(A.44)]{Besken:2019bsu} and turns out be the following 
\begin{align}\label{primary-wf}
 \bra{u} \O(z) \ket{0} = \exp \left[  2 (\mu -i\lambda) \sum_{n=1}^\infty z^n u_n\right]~. 
\end{align}
This follows from Ward identities which can be cast into a partial differential equation and solved. To make matters explicit, we need to write this expression as a linear combination of orthonormal descendants. To achieve this, we use the identity \eqref{master}
\begin{align}
	\langle u |\O(z)|0\rangle = 1+ \sum_{N=1}^{\infty} z^N \!\! \sum_{\sum jm_j =N}\!\! [2(\mu-i\lambda)]^{\sum m_j} \frac{u_1^{m_1} u_2^{m_2}  \cdots }{m_1! m_2! \cdots}~. 
\end{align}
Therefore, the evolved state \eqref{evo2} in the oscillator basis is given by 
\begin{align}
\bra{u}	e^{i\alpha t (l_1+ l_{-1})} \O(0) \ket{0} =   e^{\alpha_0h } \left[1+ \sum_{N=1}^{\infty} (\alpha_-)^N \!\! \sum_{\sum jm_j =N}\!\! [2(\mu-i\lambda)]^{\sum m_j} \frac{u_1^{m_1} u_2^{m_2}  \cdots }{m_1! m_2! \cdots}\right]~. 
\end{align}
Finally, we can normalize the oscillator monomials and write these in terms of the orthonormal descendants $\Phi_{\{m_j\}}$ of  \eqref{Phi-def}
\begin{align}\label{psi-O}
\!\!\!\Psi_{\O}(t)\equiv	\bra{u}e^{i\alpha t (l_1+ l_{-1})} \O(0) \ket{0} = e^{\alpha_0 h } \left[ 1 + \sum_{N=1}^{\infty} (\alpha_-)^N \!\! \sum_{\sum jm_j =N}\!\! \frac{[2(\mu-i\lambda)]^{\sum m_j}}{\sqrt{T_{1,m_1} T_{2,m_2} \cdots}} \Phi_{\{m_i\}}(u)\right],
\end{align}
with $T_{j,m} =(2j)^m m!$. The coefficients in the above linear combination can now be safely extracted. These are the `wavefunctions' $\varphi_{\{m_j\}}$ 
\begin{align}\label{wavefunctions}
	\varphi_{\{m_j\}}(t) =  \frac{(\alpha_-)^N}{\cosh^{2h} (\alpha t )}  \frac{[2(\mu-i\lambda)]^{\sum m_j}}{\sqrt{T_{1,m_1} T_{2,m_2} \cdots}}~,\qquad \sum_j jm_j =N~. 
\end{align}
Note that $\alpha_{-}$ also contains a time dependence -- \cref{alpha-defs}. The time-dependence is exactly the same as the SL(2,R) wavefunctions studied in \cite{Caputa:2021sib}. In contrast to \cite{Caputa:2021sib}, the normalization of the wavefunctions are different as the probabilities are now shared between all possible Virasoro descendants and not just those corresponding to $L_{-1}^n\ket{h}$.   These wavefunctions can be thought of as weights associated with the vertices of the Young's lattice.  The quantities $|\varphi_{\{m_j\}}(t) |^2$ furnish  probabilities of finding the system in a particular descendant state at time $t$
\begin{align}\label{probabilities}
	p_{\{m_j\}}(t)=|\varphi_{\{m_j\}}(t)|^2 =\frac{\tanh^{2N}(\alpha t)}{\cosh^{4h} (\alpha t )}  \frac{[4h]^{\sum m_j}}{ {2^{m_1} m_1! 4^{m_2} m_2! 6^{m_3} m_3! \cdots}}~. 
\end{align}
This forms one of the key results of our work.
The above weights are similar in spirit to the Plancherel measure, which is a well-known  probability measure over the space of integer partitions/Young diagrams. As a sanity check, we can verify that the above coefficients \eqref{wavefunctions} when summed over all descendant levels give unity -- this is carried out  in Appendix \ref{app:norms}.\footnote{If we keep aside the time-dependent part of the expression \eqref{probabilities}, the remaining factor is same as the coefficients that appear in the cycle index of symmetric group $S_N$ with mild variable redefinitions. This coincidence can be traced back to \eqref{primary-wf} which is a close cousin of the generating function of the $S_N$ cycle index. The coefficients of the cycle index are the weights of the allowed permutations of length $N$. It is no surprise that the configuration of cycles of the permutations also have a one-to-one correspondence to Young diagrams.} 

\begin{figure}[!t]
	\centering
	\includegraphics[width=0.75\linewidth]{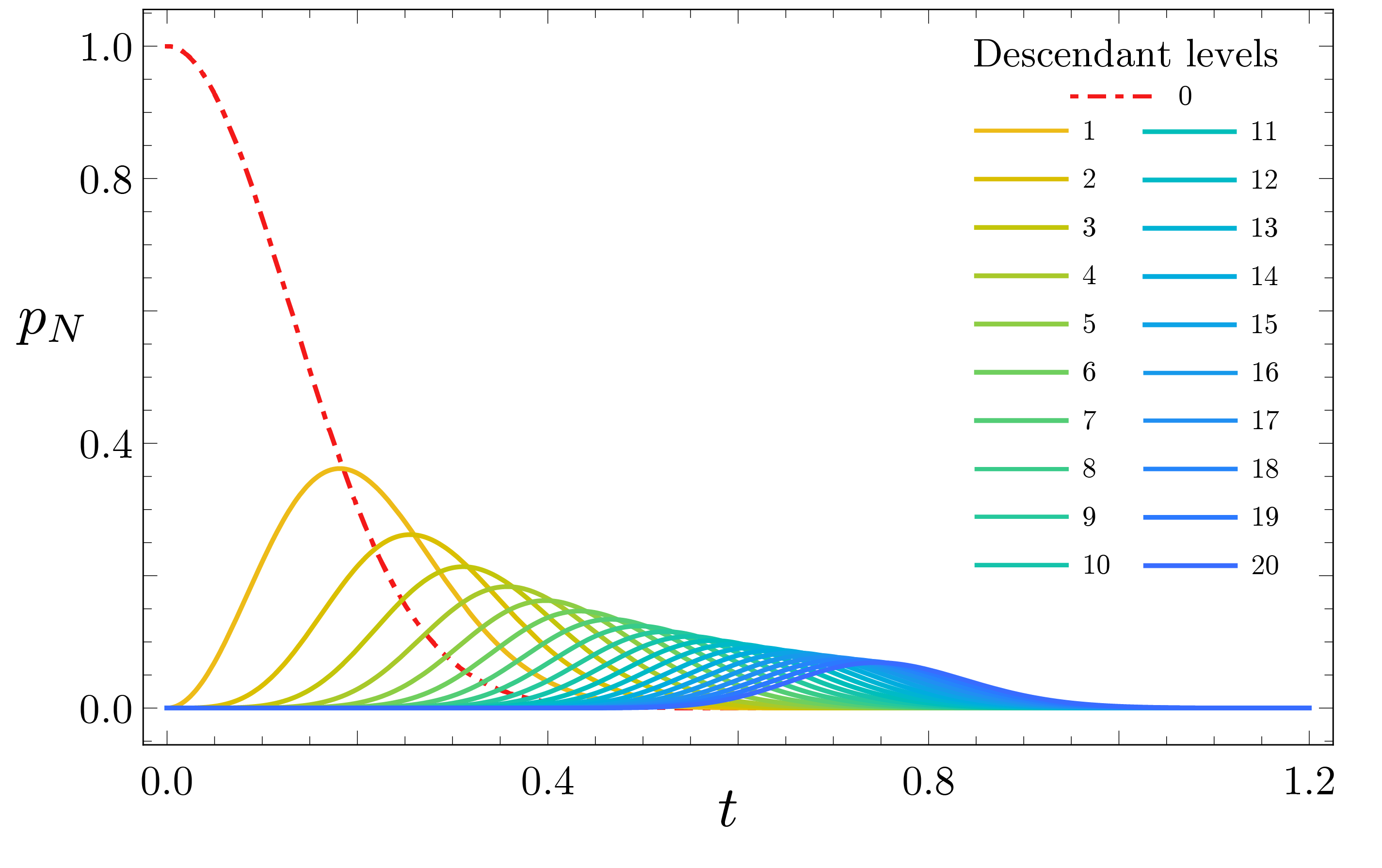}
	\caption{Plot of the probabilities $p_N(t)$,  \cref{prob-coarse},  of being in descendant levels 0 to 20 with $h=15$. The probability of remaining in the primary state ($p_\varnothing(t)$ for $N=0$) is shown by the red dashed curve. }
	\label{fig:probabilities}
\end{figure}

The exact expression for the probabilities  \eqref{probabilities} allows us to carefully examine the dynamics of the operator growth/spreading. 
We notice that at $t=0$ the probability  \eqref{probabilities} is fully concentrated on the primary state alone (or, at the tip of the lattice \cref{fig:3dyoung}) as this is our initial state. In other words, we have $p_{\varnothing}=1$, while probabilities of all other descendants vanish.  For $t>0$, all descendants acquire non-vanishing probablities (or, $p_\varnothing$ spreads out into other vertices of the lattice).   The situation is roughly analogous to the delocalization of a delta-function into a Gaussian-like form in a transport/dissipation process.  In our case, however, we have a discrete spreading along the Young's lattice. As we saw in \cref{subsec:complex-defs}, a continuum limit of the recursion relation between the wavefunction and the Lanczos coefficients can be taken and it leads to a first order wave equation \cref{flow-equation} \cite[Sec 3]{Barbon:2019wsy}. The Lanczos coefficients are local speeds at which the the initial wavefunction spreads out. Furthermore, we also see that high-level descendants become more probable at late times -- see \cref{fig:probabilities}.\footnote{From \eqref{probabilities} we have, $p_{\{m_j\}}(t\to \infty)\propto  e^{N/\xi(t)}e^{-2h\alpha t}$, with the delocalization length, $\xi(t)\sim e^{2\alpha t}$ -- see \cite[Sec V.A]{Parker:2018yvk}. Hence, the probability of low-level descendants are comparatively more suppressed than the high-level ones at late times. }

Using the expression for the individual probabilities \eqref{probabilities}, we can evaluate the net probability of reaching a specific descendant level $N$. This is given by the sum over probabilities of the descendants of that level
\begin{align}\label{prob-coarse}
	 p_{N}(t) = \sum_{\sum{j m_j}=N} p_{\{m_j\}}(t) = \frac{\tanh^{2N}(\alpha t)}{\cosh^{4h} (\alpha t )} \sum_{\sum{j m_j}=N}  \frac{[4h]^{\sum m_j}}{ {2^{m_1} m_1! 4^{m_2} m_2! \cdots}} = \frac{\Gamma (2 h+N)    }{N! \Gamma (2 h)}  \frac{\tanh^{2N}(\alpha t)}{\cosh^{4h} (\alpha t )}~.
\end{align}
Here, we have used the identity \eqref{master-prob}. A plot of these probabilities is shown in \cref{fig:probabilities}.    This probability matches with the SL(2,R) case studied in \cite[eq.\,(41)]{Caputa:2021sib} and \cite[eq.\,(25)]{Parker:2018yvk}. However, note that this agreement occurs only after coarse-graining over a given descendant level. The individual probabilities \eqref{probabilities} contain more detailed information about  the evolution. 

The auto-correlation function or return probability is given by the inner product with the primary state itself. This is nothing but the  coefficient $\varphi_{\varnothing}$ for $N=0$. It is, from \eqref{wavefunctions}
\begin{align}\label{auto-corr}
	C(t) \equiv \big(1,\Psi_{\O}(t)\big) = \frac{1}{\cosh^{2h}(\alpha t)}~. 
\end{align}
As expected, this shows an exponential decay at late times and, also, heavier primaries decay faster. 
The above auto-correlator agrees with the SL(2,R) case considered in \cite{Caputa:2021sib} and it also coincides with the auto-correlator in the inner product \eqref{wightman} upon identifying $\alpha=\pi/\beta$. The agreement with the SL(2,R) case is not surprising as the above expectation value \eqref{auto-corr} is fixed by the global symmetry alone.
\subsection{Krylov complexity}
\label{subsec:krylov}

Now that we have the exact wavefunctions \eqref{wavefunctions} at our disposal, we can evaluate the Krylov complexity. In systems without degeneracies, the Krylov complexity describes the average position in the 1d lattice of the Lanczos sequence \cite[Fig.\,1]{Parker:2018yvk}. For the case of 2d CFTs, the lattice is not one-dimensional anymore as we have the Young's lattice. The Krylov complexity should then provide a measure of the average `layer' of the Young's lattice (\cref{fig:younglattice-main}) at time $t$ (equivalently, this is the average descendant level reached after time $t$ has elapsed). Since there are degeneracies from the descendants, we need to generalize the standard definition \eqref{K-com-def}. A natural generalization should not involve additional weights for descendants of a fixed level. It is the following
\begin{align}\label{ko-1}
	K_{\O}(t) &=  \sum_{N=0}^\infty  N \sum_{ {\sum jm_j =N} } | \varphi_{\{m_j\}}(t) |^2 ~. 
\end{align}
Note that, for the specific case of 2d CFTs, the above formula of the K-complexity can also be re-expressed using expectation values of $L_0$ 
\begin{align}\label{avg-desc}
		K_{\O}(t) &=  \vev{\O(t)|L_0|\O(t)} - \vev{\O(0)|L_0|\O(0)} = \vev{\O(t)|L_0|\O(t)} -h   \equiv \vev{\hat N},
\end{align}
where, $\ket{\O(t)}$ is defined in \eqref{evo2} and the operator $\hat N \equiv L_0 - h$. It is clear from the above relation that the K-complexity measures the average descendant level or the average layer of the Young's lattice reached at time $t$.

We can plug in the wavefunctions \eqref{wavefunctions} in \eqref{ko-1} to evaluate the K-complexity 
\begin{align}\label{ko-2}
	K_{\O}(t)
	= \frac{1}{\cosh^{4h} (\alpha t )} \sum_{N=1}^\infty N |\alpha_{-}|^{2N} \sum_{\sum jm_j= N} \frac{[4(\mu^2+\lambda^2)]^{\sum m_j}}{{2^{m_1}m_1!4^{m_2}m_2!  \cdots}} ~. 
\end{align}
To evaluate the sum above we use the logarithmic derivative of the master identity \eqref{master} along with the identifications below
\begin{align}\label{masterx}
	&z\pd_z \exp\left[ \sum_{n=1}^\infty z^n y_n \right] = \sum_{N=1}^{\infty} Nz^N \!\! \sum_{\sum jm_j =N}\!\frac{y_1^{m_1} y_2^{m_2}  \cdots }{m_1! m_2! \cdots}~, \qquad  y_n = \frac{4(\mu^2+\lambda^2)}{2n} = \frac{2h}{n},~  z= |\alpha_-|^2~. 
\end{align}
Resumming the argument of the exponential and then taking the derivative yields
\begin{align}
	  z\pd_z \exp\left[ \sum_{n=1}^\infty z^n \frac{2h}{n}\right] =  \frac{2hz}{(1-z)^{2h+1}} ~. 
\end{align}
Therefore, the Krylov complexity of the operator $\O$ turns out to be
\begin{align}\label{krylov-prim}
	K_{\O}(t) &= \frac{1}{\cosh^{4h} (\alpha t )}\frac{2h|\alpha_{-}|^2 }{(1-|\alpha_{-}|^2 )^{2h+1}}  =2h\sinh^2 (\alpha t)~.  \end{align}
Once again this agrees with the SL(2,R) result  \cite{Caputa:2021sib}. 
However, note that $K_{\O}(t)$ includes contributions from all possible Virasoro descendants  and not just those of the kind $L_{-1}^n\ket{h}$ as in \cite{Caputa:2021sib}. The reason why we see an agreement with the SL(2,R) case is because the K-complexity \eqref{ko-1} involves a coarse-graining over probabilities from each descendant level -- cf.\,\eqref{prob-coarse}.  
For $t=0$, we get $K_{\O}=0$ as expected. For late times, we obtain an exponential growth of the complexity
\begin{align}\label{KO}
	K_{\O}(t\to \infty) &\approx \frac{h}{2} e^{2\alpha t}~. 
\end{align}

Let's now take a look at the fluctuations around the  value of the K-complexity  \eqref{krylov-prim}. As the K-complexity for our case has the interpretation as the average descendant level \eqref{avg-desc}, we can obtain the fluctuations through the normalized-variance or K-variance \eqref{K-variance-def} as follows
\begin{align}\label{K-variance}
\delta_{\cO}(t)^2\equiv \frac{{\rm Var}[K_\O(t)]}{K_\O(t)^2}=	\frac{\vev{\hat N^2}- \vev{\hat N}^2}{\vev{\hat N}^2} ~, \qquad \vev{\hat N^2}  =  \sum_{N=0}^\infty  N^2 \sum_{ {\sum jm_j =N} } | \varphi_{\{m_j\}}(t) |^2 ~. 
\end{align}
This can be calculated using manipulations similar to those leading to \eqref{krylov-prim} and, once again, makes crucial use of the identity \eqref{master} or, equivalently, \eqref{master-prob}. We skip the details here and provide the final result 
\begin{align}
	\delta_{\cO}(t) = \frac{\coth   (\alpha t)}{\sqrt{2h} } ~\stackrel{t\to\infty}{\approx}~ \frac{1}{\sqrt{2h} \, }~. 
\end{align}
Therefore, the fluctuations stabilize to a constant value at late times. Furthermore, we see that the fluctuations are small for heavy primaries.

Although the late-time exponential growth \eqref{KO} is expected on general physical grounds, we conclude that the total K-complexity \eqref{krylov-prim} is not sensitive enough to distinguish between the SL(2,R) and Virasoro cases for simple primary operators in CFTs on an infinite line.\footnote{The case of finite size and temperature, which is a torus, would not be universal.} However, using our present formalism, we can consider subsets of vertices in the Young's lattice (or subsets of descendants) and find the K-complexities associated with them. This provides fine-grained information specific to the Virasoro case. We turn to this in what follows. 

\subsubsection*{Single row diagrams}
\begin{figure}
	\centering
	\includegraphics[width=\textwidth]{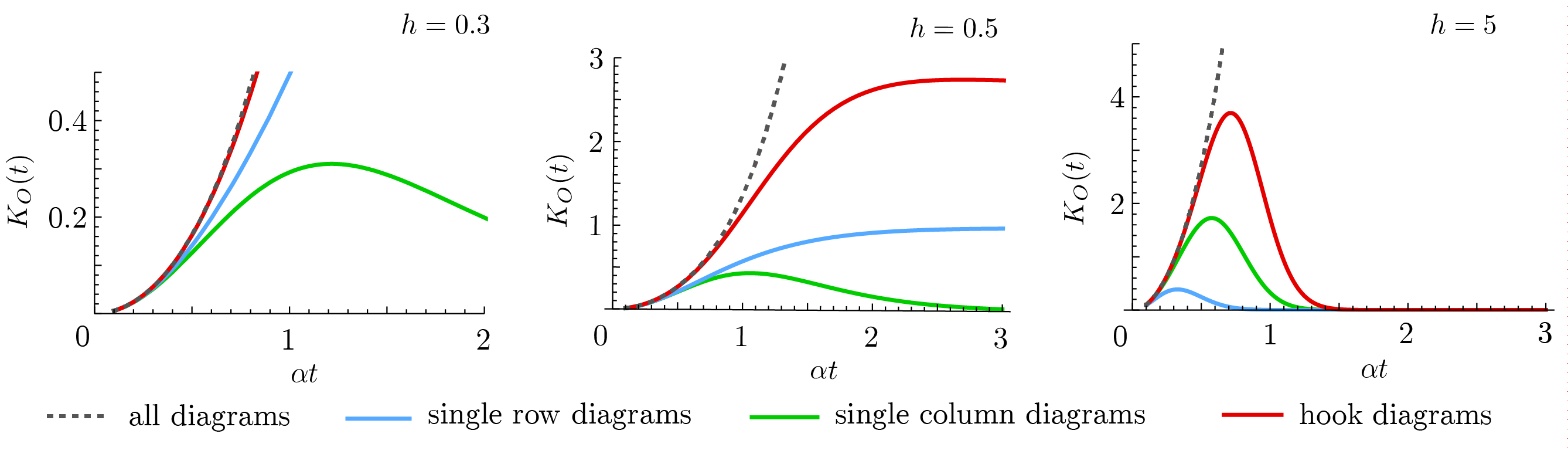}
	\caption{Plots of K-complexities.}
	\label{fig:kplots}
\end{figure}

As our first case, we consider the K-complexity for the descendants that lie at the rightmost edge of the Young's lattice, \cref{fig:younglattice-main} (left). These are single row Young diagrams that correspond to states of the kind $\ket{N^1}$ or simply proportional to $u_N$ in the oscillator variables. We found in the previous section that the Lanczos coefficients along this path have maximal growth. One might naively expect that this gives dominant contribution to the total K-complexity at late times. However, this is not the case. The K-complexity of this subset of states is 
\begin{align}\label{K-row}
	K^-_{\O}(t) &=  \sum_{N=1}^\infty  N  | \varphi_{\{0,0,0,\cdots,1_N\}}(t) |^2 = \frac{2h}{\cosh^{4h} (\alpha t )}\sum_{N= {1}}^\infty   |\alpha_{-}|^{2N}    = \frac{2h \sinh^2(\alpha t)}{\cosh^{4h} (\alpha t )}  ~. 
\end{align} 
For $h>1/2$ we have an initial rise followed by a decay, for $h<1/2$, we have a monotonically growing behaviour, while for $h=1/2$ we have an initial rise followed by a saturation at late times -- see the blue curves in \cref{fig:kplots}. This shows that the descendants of this kind are not the dominant contribution to the net K-complexity at late times. The exponential rise of \eqref{KO}  emerges as a collective effect from all diagrams of the lattice.

\subsubsection*{Single column diagrams}

We now consider the K-complexity for the descendants of the kind $\ket{1^N}$ that correspond to single column Young diagrams. In terms of oscillators these are monomials of the form $u_1^N$.  
As demonstrated in the previous section, the Lanczos coefficients connecting these states have $\sqrt{N}$ growth. The K-complexity is
\begin{align}
	K^{|}_{\O} (t)&=  \sum_{N=0}^\infty  N  | \varphi_{\{N,0,0,\cdots\}}(t) |^2 = \frac{1}{\cosh^{4h} (\alpha t )} \sum_{N=1}^\infty  {\left(2h|\alpha_{-}|^2 \right)^{N} \over (N-1)! } 
	= \frac{2h\sinh^2(\alpha t)}{\cosh^{4h+2} (\alpha t )}e^{2h\tanh^2(\alpha t)}~. 
\end{align} 
For all $h>0$, the profile invariably shows an initial exponential rise, followed by reaching a maxima and then decaying at later times -- the green curves in \cref{fig:kplots}.

\subsubsection*{Hook diagrams}

Finally, we consider hook diagrams that are an intermediate class of states that lie in between single row/column diagrams. These are states of the kind $\ket{1^{N-r} r^1} \leadsto u_1^{N-r}u_r^1$. For $r=1$ we have   single column diagrams, while for $r=N$ we have  single row diagrams. The K-complexity can be evaluated as follows. 
 The first term in square brackets below is the $r=1$ term, and is treated separately
\begin{align}\label{K'''}
	K\ulcorner\!\!_{\O}(t) &=  \sum_{N=1}^\infty  N \left[ | \varphi_{\{N,0,0,\cdots,0\cdots,0\}}(t) |^2 + \sum_{r=2}^N | \varphi_{\{N-r,0,0,\cdots,0,1_r,0\cdots,0\}}(t) |^2 \right] \nn \\
	&=  K^|_{\O}(t) +   \frac{1}{\cosh^{4h} (\alpha t )} \sum_{N=1}^\infty  N |\alpha_{-}|^{2N} 
	\sum_{r=2}^{N}  
	\frac{(4h)^{N-r+1}}{2^{N-r} (N-r)! (2r)^1 1! }~. 
\end{align} 
Note that the $r=N$ term gives the $K^-_{\O}(t)$ contribution, \cref{K-row}. 
The summations in the second term can be performed by interchanging the order. We skip the details here and provide the final result
\begin{align}
  	K\ulcorner\!\!_{\O}(t)={2 h \sinh ^2(\alpha  t) \over \cosh ^{4 h}(\alpha  t)} e^{2 h \,\tanh ^2(\alpha  t)} \left[2 h\, \text{sech}^4(\alpha  t)+2 h\, \text{sech}^2(\alpha  t) \left(\log \left|\text{cosh}^2(\alpha  t)\right|-1\right)+1\right]   ~. 
\end{align}
It is obvious that for all times, $K\ulcorner\!\!_{\O}(t)  \geq K^-_{\O}(t) + K^|_{\O}(t)$. The temporal profile is similar to what we saw for the single column diagrams -- monotonically increasing for $h<1/2$, initial rise and saturation for $h=1/2$ and rise followed by decay for $h>1/2$. This is shown in the red curves in \cref{fig:kplots}. 

We contrast the K-complexities of single-row/column/hook diagrams with the total K-complexity in \cref{fig:kplots}. It is clear that the net exponential growth of the K-complexity is collective effect from all paths of the lattice. Furthermore, the hook diagrams and single row/diagrams are highly atypical at high descendant levels \cite{Datta:2019jeo}. Therefore, the relative contribution from these atypical states dampens out with time and the major contribution originates from the  large number of typical descendant states. Let us now see how the K-complexity \eqref{krylov-prim} can be interpreted geometrically. 

\subsubsection*{Information metric and volume}
The K-complexity for primary operators can be associated with the volume in the information metric; see \cite{Caputa:2021sib} for the SL(2,R) case. We shall now demonstrate this relation using the evolved state \eqref{psi-O}. 

We consider a slightly generalized two-parameter family of normalized coherent states labelled by complex variable $z$, analogous to \eqref{psi-O}
\bea\label{coh}
\ket{z,h}= (1-z\bar{z})^{h}\left[ 1 + \sum_{N=1}^{\infty} z^N \!\! \sum_{\sum jm_j =N}\!\! \frac{[2(\mu-i\lambda)]^{\sum m_j}}{\sqrt{T_{1,m_1} T_{2,m_2} \cdots}} \Phi_{\{m_i\}}(u)\right]~.
\eea
The evolved state \eqref{psi-O} can be thought of as a trajectory through this space of states (phase-space), given by the parametrization in \eqref{TimeDepPar} -- with $z=re^{i\phi}$.  We can associate a natural information metric or distance measure for these general coherent states. This is the Fubini-Study distance
\be\label{FSM}
ds^2=\langle dz,h|dz,h\rangle-\langle dz,h|z,h\rangle\langle z,h|dz,h\rangle~.
\ee
From the state \eqref{coh} we derive
\be
\ket{dz,h}=-\frac{h(zd\bar{z}+\bar{z}dz)}{1-z\bar{z}}\ket{z,h}+\frac{dz}{z}(1-z\bar{z})^h\left[\sum_{N=1}^{\infty} Nz^N \!\! \sum_{\sum jm_j =N}\!\! \frac{[2(\mu-i\lambda)]^{\sum m_j}}{\sqrt{T_{1,m_1} T_{2,m_2} \cdots}} \Phi_{\{m_i\}}(u)\right].
\ee
Then, using the master identity \eqref{master} and its logarithmic derivatives we obtain
\be
\langle z,h|dz,h\rangle=h\frac{\bar{z}dz-zd\bar{z}}{1-z\bar{z}}=-\langle dz,h|z,h\rangle~. 
\ee
The first term in \eqref{FSM}, $\langle dz,h|dz,h\rangle$, can also be evaluated analogously. We skip the detailed calculations here as the manipulations involved are essentially the same as those performed while obtaining the K-complexity earlier in this section. The final result is that the metric becomes the two-dimensional hyperbolic disc
\bea\label{FSPrim}
ds^2=\frac{2hdzd\bar{z}}{(1-z\bar{z})^2}~.
\eea
The volume of the disc contained within  the surface, $r=\alpha t$, becomes
\be\label{volume}
V(t)=2\pi h\sinh^2 (\alpha t)=\pi K_{\O}(t)~.
\ee
This shows the claim that the Krylov complexity \eqref{krylov-prim} is proportional to the volume in the Fubini-Study metric. 

There's one aspect of the above result we would like to emphasize. In contrast to the SL(2,R) case \cite{Caputa:2021sib},  the integral for the inner products \eqref{InnerproductVir} is not over the unit disk and its measure has nothing to with SL(2,R). It is interesting to note that, although this is the case, we still get the two-dimensional hyperbolic geometry \eqref{FSPrim} in the complex coordinate $z$. This can be seen as a consequence of the fact that the information geometry  is associated with a Liouvillian  and a quantum circuit constructed purely from SL(2,R) generators -- \eqref{L-algo} and \eqref{L-algo-2}.  From the perspective of the Young's lattice, the local action of an addition or removal of a single box corresponds to an SL(2,R) action. Hence, the effective geometry associated with the edges of the graph has this group as its isometries. 

\subsection{Renyi entropies}
\label{subsec:renyi}
We can extract further properties of the probability distribution corresponding to \eqref{wavefunctions} by studying entropic measures associated with it. 
The K-entropy is defined through the von Neumann entropy of the probabilities \eqref{probabilities}
\begin{align}\label{Kent}
	S_K(t)&=  -\sum_{N=0}^\infty    \sum_{ {\sum jm_j =N} } | \varphi_{\{m_j\}}(t) |^2 \log | \varphi_{\{m_j\}}(t) |^2 ~. 
\end{align}
It turns out that this quantity is rather difficult to simplify for the $\varphi_{\{m_j\}}$'s in \eqref{wavefunctions}. 
A closely related quantity is the Renyi K-entropy, defined through the moments of the probability distribution. Using the wavefunction \eqref{wavefunctions} we get
\begin{align}\label{K-Renyi-def}
	S^n_K&=  \frac{1}{1-n}\log\sum_{N=0}^\infty  \sum_{ {\sum jm_j =N} }    | \varphi_{\{m_j\}}(t) |^{2n}  \\
	&=  \frac{1}{1-n}\log\left[\frac{1}{\cosh^{4nh} (\alpha t )} \sum_{N=0}^\infty   |\alpha_{-}|^{2Nn} \sum_{\{m_j\}} \frac{[4h]^{n\sum m_j}}{{2^{nm_1}(m_1!)^n 4^{nm_2}(m_2!)^n 6^{nm_3}(m_3!)^n  \cdots}}\right]~. \nn 
\end{align}
To evaluate the above sum we use the generalized master identity \eqref{master2} that involves hypergeometric functions
\begin{align}
	 \prod_{p=1}^\infty {}_0F_{n-1}(1,1,\cdots,1|(z^p y_p)^n)  = 1+ \sum_{N=1}^{\infty} z^{Nn} \!\! \sum_{\sum jm_j =N}\!\! \frac{y_1^{nm_1} y_2^{nm_2}  \cdots }{(m_1!)^n (m_2!)^n \cdots} ~. 
\end{align}
For our specific case, in \eqref{K-Renyi-def}, we have the identifications
\begin{align}
	y_p = \frac{2h}{p}, \qquad z= |\alpha_-|^2~. 
\end{align}
The K-Renyi entropy then becomes
\begin{align}\label{SK}
	S_K^n
	&=  \frac{1}{1-n}\left[ \sum_{p=1}^\infty \log {}_0F_{n-1}\left(1,1,\cdots,1\bigg|
	\frac{(2h)^n}{p^n} \tanh^{2pn}(\alpha t) 
	\right) - 4nh \log {\cosh  (\alpha t )}\right]  ~. 
\end{align}
It can be seen  that the expression has the expected $0/0$ form in the $n\to 1$ limit, upon using ${}_0F_0(x)=e^x$. However, the analytic continuation to the K-entropy \eqref{Kent} is not straightforward as $n$ is a discrete index in the generalized hypergeometric function appearing above.\footnote{That said, partial progress towards finding the K-entropy can be made. The expression \eqref{SK} can be written using the Le Roy function, \eqref{leRoy-def}, that allows an analytic continuation beyond integers.
We can then focus on the regime in which the operator, $O$, is a heavy primary, $h\to \infty$. The asymptotics of the Le Roy function, \eqref{leRoy-asymp}, can be used provided we truncate the sum over to some $p_{\rm max} \ll h$. The contribution to the K-entropy for this subset of terms (in the sum over $p$) can then be obtained upon taking the $n\to 1$ limit.
}

We note here that the K-Renyi entropy \eqref{SK} is different from the SL(2,R) case considered in \cite{Caputa:2021sib}. This is in contrast to the K-complexity \eqref{krylov-prim}, which is the same for both Virasoro and SL(2,R) cases. The reason why K-Renyi entropies are more sensitive is that, unlike the K-complexity defined in \eqref{ko-1}, they do not have a coarse-graining (or averaging) over each descendant level. Therefore, $S^n_K$ is a fine-grained probe that depends more sensitively on the detailed, individual probabilities \eqref{probabilities} of reaching the descendant states.  

\section{Growth of the stress tensor}
\label{sec:stress}

The techniques developed above can also be utilized to study the evolution of the (holomorphic) stress tensor $T(z)$ and the K-complexity associated with it. From an AdS/CFT perspective, the growth of the stress tensor is something that belongs to the pure gravity sector and is expected to capture some universal properties of the dual CFT. Furthermore, it is valuable to investigate the operator growth of conserved currents of the KdV hierarchy under the protocol \eqref{L-algo}, and the stress tensor is simplest one in this regard.

The stress tensor is also an interesting example since its modes are genuinely associated with the underlying Virasoro symmetry of the 2d CFT \eqref{Tz}. Moreover, it is not a CFT primary and has a non-trivial transformation under conformal maps $z\mapsto f(z)$
\be
T(z)=f'(z)^2T(f(z))+\frac{c}{12}\{f(z),z\},
\ee
with the second term being the Schwarzian derivative. The two-point function in a CFT on an infinite line at inverse temperature $\beta$ is then
\be
\langle T(w_1)T(w_2)\rangle_\beta=\frac{c}{2}\left(\frac{\beta}{\pi}\sinh\left(\frac{\pi w_{12}}{\beta}\right)\right)^{-4}+\left(\frac{\pi^2c}{6\beta^2}\right)^2.
\ee
Consequently, the real time auto-correlator \eqref{AutoCor} that is obtained by setting $w_2=0$ and $w_1=i\left(\beta/2+it\right)$ becomes
\be
C(t)=\langle T(t)T(0)\rangle_\beta=\frac{c}{2}\left(\frac{\beta}{\pi}\cosh\left(\frac{\pi t}{\beta}\right)\right)^{-4}+\left(\frac{\pi^2c}{6\beta^2}\right)^2.
\ee
Naively, one may expect that, in the light of the standard Lanczos algorithm based on moments of $C(t)$, operator growth of the stress tensor should resemble that of a primary with dimension $h=2$. On the other hand, the fact the stress tensor is a quasiprimary and  the details of Virasoro symmetry should also play a distinct role. We reconcile this tension below and indeed confirm the naive expectation at the level of Krylov complexity, however, using the justified technology of the oscillator realization for the full Virasoro symmetry. 
 
Similar to the primary case, we can evaluate the evolution of the stress tensor analytically. The coherent state is again given by a displacement operator acting on the CFT state corresponding to $T(0)$, \ie the descendant $L_{-2}\ket{0}$
\be
e^{(\xi L_{-1} -\bar \xi L_{1})} T(0) \ket{0} =e^{\alpha_- L_{-1}} e^{\alpha_0 L_{0}} e^{\alpha_+ L_{1}} L_{-2} \ket{0}=e^{2\alpha_0 }e^{\alpha_- L_{-1}} L_{-2} \ket{0}, 
\ee
where in the second step we used the BCH formula and the third step follows from the fact that the $L_0$ and $L_1$ eigenvalues of the state $L_{-2}\ket{0}$ are  $2$ and 0 respectively. The time dependence is the same as in the primary case \eqref{TimeDepPar}. Moreover, the above state can be written as
\be
e^{2\alpha_0 }e^{\alpha_- L_{-1}} L_{-2}e^{-\alpha_- L_{-1}}  \ket{0}=e^{2\alpha_0 }\left(L_{-2}+\alpha_-[L_{-1},L_{-2}]+\frac{\alpha_-^2}{2}[L_{-1},L_{-3}]+...\right)\ket{0}~. 
\ee
Using the Virasoro algebra this becomes
\be
e^{2\alpha_0 }T(\alpha_-)\ket{0}~,\qquad \alpha_-=i\tanh(\alpha t).
\ee
Finally, we can appropriately normalize the state by the norm of the initial state, $L_{-2}\ket{0}$. This is fixed, once again, by the Virasoro algebra and leads to
\begin{align}\label{ell-}
	\ket{\Psi_T}=\sqrt{2\over c}e^{2\alpha_0} 	T(\alpha_{-})\ket{0} =\sqrt{2\over c} e^{2\alpha_0}  \sum_{m=2}^{\infty} (\alpha_-)^{m-2} L_{-m} \ket{0} ~. 
\end{align}
We can now write the above quantity in the oscillator basis, using the differential operator realization of the Virasoro generators from \eqref{vira}. In addition, we need to set $\lambda=i\mu$ as we are dealing with descendants of the identity -- due to technicalities with conjugation \eqref{conjugation}, we will keep $\mu$ and $\lambda$ intact for the time being and impose $\lambda^2 = - \mu^2$ while nearing the end of our calculations. In the oscillator basis, we have
\begin{align}\label{l2osc}
	\Psi_T(u)\equiv\langle u\ket{\Psi_T}&=\sqrt{2\over c} e^{2\alpha_0}  \sum_{m=2}^{\infty} (\alpha_-)^{m-2} \bra{u} l_{-m} \ket{0} ~, \\
	&= -\sqrt{\frac{2}{c}}e^{2\alpha_0}  \sum_{m=2}^{\infty} (\alpha_-)^{m-2} \left[ \sum_{j=1}^{m-1} j(m-j)u_ju_{m-j} - 2m(\mu m-i\lambda)u_m\right]~. \nn 
\end{align}
Note that the monomials appearing above are not normalized to unity. The non-vanishing inner products are the following 
\begin{align}\label{uu-norms-0}
	(u_ju_{m-j},u_pu_{m-p}) \equiv {\cal M}_{j,p}
	=\begin{cases}
		\frac{1}{4j(m-j)} \quad \text{if }j=p \\
		\frac{1}{4j(m-j)} \quad \text{if }m-j=p \\ 
		\frac{2}{m^2}\qquad\quad  \!\text{if }j=p=m/2 \text{ for even } m
	\end{cases}\hspace{-0.5cm},~ (u_m,u_m)  = \frac{1}{2m}\,.
\end{align}

Although it is an obvious fact, it is instructive to verify that the full expression \eqref{l2osc} above indeed has a unit norm; this is checked in Appendix \ref{app:norms}. Furthermore, the evaluation of K-complexity proceeds along the same lines as the verification of the norm. The expression for K-complexity is (analogous to \cref{TT-norm})
\begin{align}\label{K-stress}
	K_T(t)=&\frac{2}{c}e^{4\alpha_0}  \sum_{m=2}^{\infty} (m-2) |\alpha_-|^{2m-4} \left[  \sum_{j,p=1}^{m-1}  j^2(m-j)^2 {\cal M}_{j,p}  + {2m}(\mu^2 m^2{+\lambda^2}) \right] ~. 
\end{align}
The offset $(m-2)$ is present since  our initial state is $L_{-2}\ket{0}$, \ie it is a descendant at level 2 and the Liouvillian evolution takes us to descendants of levels $\geq 2$. We can  set, $\lambda^2=-\mu^2= -(c-1)/24$, at this stage.
For both odd and even $m$, the sum over $j,p$ in \eqref{K-stress} evaluates to 
\begin{align}
	&	 \sum_{j,p=1}^{m-1}  j^2(m-j)^2 {\cal M}_{j,p}  + {2m}\mu^2( m^2-1 )  =
	\frac{c}{12} m(m^2 - 1) ~,
\end{align}
see \cref{m-odd-sum,m-even-sum} for details. The K-complexity then evaluates to, after using \eqref{alpha-defs},  
\begin{align}\label{KT}
	K_T(t)   = 4 \sinh^2 (\alpha t)~.
\end{align}
This has the same form as that of \eqref{krylov-prim} upon setting $h=2$ and grows exponentially at late times. Moreover, for the same value of $h$ this expression matches with the expression for the SL(2,R) primaries \cite{Caputa:2021sib}. We can also obtain the fluctuations or the K-variance,  just like the case of primaries  \eqref{K-variance},  and it turns out to be 
$
	\delta_T(t) =  {\coth (\alpha t)}/{2}. 
$

\subsubsection*{Information metric for the stress tensor}
Analogous to the case of primary operators, we can evaluate information metric \eqref{FSM} for the stress tensor. Once again, we consider a two-parameter family of generalized coherent states
\be\label{CSTz}
\ket{z,c}\equiv-\sqrt{\frac{2}{c}}(1-z\bar{z})^2\sum^\infty_{m=2}z^{m-2}\left[\sum^{m-1}_{j=1}j(m-j)u_ju_{m-j}-2m(\mu m-i\lambda)u_m\right],
\ee
such that the state representing the evolution of the stress tensor, \eqref{l2osc}, in the Krylov space corresponds to the trajectory \eqref{TimeDepPar} in this family. 

To derive the information metric \eqref{FSM} it is useful to write the coherent state \eqref{CSTz} compactly as
\bea\label{ZC}
\ket{z,c}=-\sqrt{\frac{2}{c}}(1-z\bar{z})^2\sum^\infty_{m=2}z^{m-2}\tilde{\Psi}_m(u),
\eea
where, $\tilde{\Psi}_m(u)$, is the quantity in square brackets in \eqref{CSTz}. The norm of the wavefunction $\tilde{\Psi}_m(u)$ (see Appendix \ref{app:norms}) is
\bea
(\tilde{\Psi}_m(u),\tilde{\Psi}_{m'}(u))=\delta_{m,m'}\frac{c}{12}m(m^2-1)~.
\eea
Proceeding further, we can  compute the derivative of \eqref{ZC}
\bea
\ket{dz,c}&=&-\frac{2(zd\bar{z}+\bar{z}dz)}{1-z\bar{z}}\ket{z,c}-\frac{dz}{z}\sqrt{\frac{2}{c}}(1-z\bar{z})^2\sum^\infty_{m=2}(m-2)z^{m-2}\tilde{\Psi}_m(u)~.
\eea
Finally, evaluating the inner products as required by the definition of the Fubini-Study metric \eqref{FSM} yields the following 
\bea
ds^2=\frac{4dzd\bar{z}}{(1-z\bar{z})^2}~.
\eea
This is the same hyperbolic geometry  that we obtained for the primary operators, eq.\,\eqref{FSPrim}, but with $h=2$. As before, this is clearly consistent with the interpretation of $K_T(t)$, in \eqref{KT},  as the volume in this metric -- see \cref{volume}. The universality of the result also confirms that the information geometry is strongly tied with the quantum circuit of the Liouvillian (or the evolution protocol) and mildly depends on the reference states.

\section{Discussions}
\label{sec:discuss}

In this paper we undertook a detailed investigation into operator growth in  2d irrational CFTs. Our focus was on the evolution of a single primary operator and we uncovered universal properties of the Liouvillian evolution that are fixed by Virasoro symmetry.
We found that Lanczos algorithm for 2d CFTs generalizes in a natural manner. Owing to the degeneracies present due to descendants, the Liouvillian matrix acquires a block tridiagonal structure. We also found that the Lanczos coefficients crucially depend on the details of the descendant states and a subset of them does saturate the upper bound of linear growth conjectured by \cite{Parker:2018yvk}. Irrational 2d CFTs, thereby, provide one of the first field-theoretic examples for fastest growth of Lanczos coefficients. We also arrived at a closed form for the evolved wavefunction using which we analyzed the dynamics of operator spreading.  All of these features have a very clear interpretation as paths spreading on the Young's lattice. We noticed that the Krylov complexity of the primary operator and the stress tensor takes the same universal form as the SL(2,R)
case and, in both cases, has an interpretation as volume in the information geometry. However, this universality is a double-edged sword since it can be interpreted as the K-complexity not being sensitive enough.  A close inspection reveals that the Lanczos matrices, $b_{\{m_j\}\to \{r_k\}}$, and the wavefunctions, $\varphi_{\{m_j\}}$, clearly contain more fine-grained information than the quantities they get repackaged in. This was seen explicitly when we considered K-complexities of a subset of descendants of a specific kind -- the temporal profiles turned out to be different depending on the value of the conformal dimension. This generalized K-complexity defined for subclasses of vertices  on the Young lattice is one of our main results. In addition, the Renyi entropies turn out to be different from the SL(2,R) case and capture more information regarding the Virasoro case.

One of our motivations to consider operator growth in $c>1$ CFTs arose from the AdS$_3$/CFT$_2$ correspondence. We now comment on possible implications of our result for the bulk dual. The inner product of the Wightman correlator \eqref{wightman} corresponds to that of a thermofield double state. This has the standard interpretation in terms of the two-sided eternal black hole in AdS$_3$. If the primary operator is reasonably heavy, $h\sim O(c)$, we can approximate it by a massive scalar particle in the bulk. The Liouvillian evolution then corresponds to the evolution of the particle's wavefunction. The probability of being in a high-level descendant state increases with time and the gravitational analogue of this phenomenon is that the particle gets dressed with several gravitons during the evolution (recall that the stress tensor corresponds to the graviton in the bulk). Roughly speaking, the Lanczos matrices provide transition amplitudes for absorption or emission of single graviton. On the other hand, the wavefunctions, $\varphi_{\{m_j\}}$, encode probabilities for being in a particular configuration of the scalar+gravitons. Note that the auto-correlator \eqref{auto-corr} can be realized in the bulk through a geodesic or Witten diagram connecting two $\O$ insertions, separated by a time-interval $t$, on either sides of the eternal black hole. It is then conceivable that information about the wavefunctions $\varphi_{\{m_j\}}$ can be obtained through graviton loop corrections of this Witten diagram.  It would be worthwhile to sharpen this picture quantitatively from the bulk perspective and find a precise correspondence with the measures considered in this work, such as the K-complexity and Renyi entropies. The correspondence with the Young's lattice can perhaps be leveraged further to translate geometrical measures of the graph to bulk quantities. On the field theory side, it would be of value to generalize the analysis of this work to understand how operator-products of primaries grow under the Liouvillian. This is a scenario which will be sensitive to the spectrum and OPE coefficients of the CFT; this implies that some general expectations from ETH and holographic CFTs will play a role.

A natural extension of our analysis is to consider other families of 2d CFTs. For instance, minimal models have null-states and the standard correspondence with the Young's diagrams/lattice requires appropriate modifications. It would be desirable to employ the Coulomb gas formalism for rational CFTs to carry out an analysis similar to the one presented here. One might expect that the maximal growth for the Lanczos coefficients observed for the irrational case would be absent for minimal models. Another class of theories which are interesting to investigate are those with extended chiral algebras -- such as, superconformal theories or $\cW_N$ CFTs containing  higher spin conserved currents. In particular, irrational $\cW_N$ theories are known to violate the bound on the Lyapunov exponent \cite{Perlmutter:2016pkf}.  It would, therefore, be desirable to extract the Lanczos coefficients and verify whether the conjectured maximal bound of \cite{Parker:2018yvk} is still obeyed.  An analogous oscillator formalism exists for the $\cW_3$ algebra \cite{Pope:1991it} which can be suitably utilized in this direction. Furthermore, the descendants of irrational $\cW_N$ theories  have a one-to-one  correspondence with $(N-1)$-coloured partitions. These can be labeled by multiple-layered Young diagrams, or equivalently by plane partitions of restricted height \cite{Prochazka:2015deb}. 
Finally, with the goal of understanding string theory in AdS$_3$, it would be worthwhile  to decipher how operator growth works in symmetric orbifold theories. The twisted sectors of the symmetric orbifold also have a one-to-one correspondence to integer partitions and Young diagrams \cite{Balasubramanian:2005qu}, a relationship that can serve as a bridging point to the analysis presented here. In the same vein, an orthonormal basis for 1/2-BPS operators in ${\cal N}=4$ super-Yang-Mills are labelled by representations of $U(N)$. This, in turn, has a one-to-one correspondence to representations of the symmetric group $S_N$ and, therefore, to Young tableaux \cite{Corley:2001zk,Balasubramanian:2005mg,deMelloKoch:2007rqf}. Just like the orthogonal basis of oscillator monomials used in this work, the orthogonal 1/2-BPS states in ${\cal N}=4$ SYM can be written using Schur polynomials. Furthermore, there are close parallels in the mathematical language to describe the dual LLM geometries \cite{Lin:2004nb} and the evolved states for 2d CFT primaries used in this work -- see \eg  \cite[Sec 3 \& 4]{Berenstein:2017abm} and \cite{Lin:2021qso}, as well as related work \cite{Simon:2018laf}.  Therefore, it would be tantalizing to develop an analogous Young lattice description underlying  operator growth for these BPS states and uncover what lies in common with the 2d CFT setup considered here.

\section*{Acknowledgments}
It's a pleasure to thank Alex Belin, Mert Besken, Justin David, Dongsheng Ge, Shajid Haque, Sinong Liu, Vishnu Jejjala, Per Kraus,  Robert de Mello Koch, Javier Magan, Dimitrios Patramanis, Tomas Prochazka, Gabor Sarosi and Joan Simon for fruitful discussions/comments on the draft. We also thank Christian Gaetz, Sam Hopkins and Richard Stanley for sharing their wisdom on the Young's lattice by answering our questions on MathOverflow. S.D.~thanks Mert Besken and Per Kraus for collaboration on related matters in the past. The work of P.C. is supported by NAWA “Polish Returns 2019” and NCN Sonata Bis 9 grants.

\appendix

\section{Summation/product identities}
\label{sec:master-id}

The following  identity is used several times in the main text. 
\begin{align}\label{master}
 &\exp\left[ \sum_{n=1}^\infty z^n y_n \right] = 1+ \sum_{N=1}^{\infty} z^N \!\! \sum_{\sum jm_j =N}\!  \frac{y_1^{m_1} y_2^{m_2}  \cdots }{m_1! m_2! \cdots}  ~. 
\end{align}
The proof of this is fairly straightforward. We start with the LHS, which can be written as an infinite product of exponentials and then expand each of the exponentials. 
\begin{align}
		\prod_{p=1}^\infty e^{z^py_p} = 	\prod_{p=1}^\infty \left[\sum_{m=0}^\infty \frac{z^{mp}y_p^m}{m!}\right] = \sum_{N=0}^{\infty} z^N \!  \sum_{\sum pm_p =N}\! \frac{y_1^{m_1} y_2^{m_2}  \cdots }{m_1! m_2! \cdots}  ~. 
\end{align}
In the final step, we have reorganized the sum by collecting powers of $z$. The powers of $z$ appear in the form of integer partition decompositions, $\sum_{p=1}^{\infty} pm_p = N$. This proves the identity \eqref{master}. 

A special case of the above identity is 
\begin{align}\label{master-prob}
	\sum_{N=0}^\infty  x^{N}	
	\sum_{{\sum j m_j =N }} \frac{(4h)^{\sum_j m_j}}{2^{m_1}m_1! 4^{m_2}m_2!\cdots }~= ~ \frac{1}{(1-x)^{2h}} ~=~ \sum_{M=0}^\infty \frac{\Gamma (2 h+M)    }{M! \Gamma (2 h)} x^M ~.
\end{align}
where, the identification $y_p = 4h/2p$ has been used in \eqref{master}. The second relation above is simply the Taylor expansion of $(1-x)^{-2h}$ around $x=0$. 

While evaluating the Renyi entropies in Section \ref{subsec:renyi}, we use a generalization of the identity \eqref{master} that involves   hypergeometric functions
\begin{align}\label{master2}
	\prod_{p=1}^\infty {}_0F_{n-1}(1,1,\cdots,1|(z^py_p)^n)  = 1+ \sum_{N=1}^{\infty} z^{Nn} \!\! \sum_{\sum pm_p =N}\!  \frac{y_1^{nm_1} y_2^{nm_2}  \cdots }{(m_1!)^n (m_2!)^n \cdots} ~. 
\end{align}
The proof of this is similar to the one above. We need to use the Taylor expansion of the hypergeometric function 
\begin{align}\label{hyp}
		{}_0F_{n-1}(1,1,\cdots,1|x^n)=
	\sum _{m=0}^{\infty } \frac{x^{nm}}{(m!)^n}  ~. 
\end{align}
Using the above expansion in the LHS of \eqref{master2} and collecting powers of $z$ as before, we arrive at the RHS of \eqref{master2} which has the summation representation.  

The function \eqref{hyp}, can actually be defined for values of any positive real value of $n$. It is often referred to as the Le Roy function. This is potentially useful for analytic continuations while studying Renyi entropies. The Le Roy function is defined as 
\begin{align}\label{leRoy-def}
	F_\rho (x) = \sum_{j=0}^{\infty} \left(x^j \over j!\right)^\rho, \qquad \forall~ 0< \rho \leq \infty~. 
\end{align}
The large $x$ asymptotics of this function is known to be \cite[pg.\,307-8]{olver1997asymptotics}
\begin{align}\label{leRoy-asymp}
	F_\rho (x) \approx \frac{e^{\rho x}}{\rho^{1/2}(2\pi x)^{(\rho-1)/2}} \left[1+ O(x^{-1})\right]~, \qquad 0< \rho \leq 4~. 
\end{align}

\section{Normalization of wavefunctions}
\label{app:norms}
\subsubsection*{Primaries}

We found that the evolved `wavefunctions', $\varphi_{\{m_i\}}(t)$, of the primary operator are given by 
\begin{align}
 	\varphi_{\{m_i\}}(t) =  \frac{(\alpha_-)^N}{\cosh^{2h} (\alpha t )}  \frac{[2(\mu-i\lambda)]^{\sum m_j}}{\sqrt{T_{1,m_1} T_{2,m_2} \cdots}}~,~\qquad \sum_j jm_j =N~.  
\end{align}
with, $T_{j,m} =(2j)^m m!$.
It's worthwhile to verify whether these are properly normalized. 
\begin{align}
	\sum_{N=0}^\infty \sum_{\{m_j\}} |\varphi_{\{m_j\}}|^2 &= \frac{1}{\cosh^{4h}(\alpha t)} \sum_{N=0}^\infty |\alpha_-|^{2N} \sum_{\{m_j\}} \frac{[4(\mu^2+\lambda^2)]^{\sum_j m_j}}{T_{1,m_1}T_{2,m_2}\cdots }  \\
	&	 = \frac{1}{\cosh^{2h}(\alpha t)} \frac{1}{(1-|\alpha_{-}|^2 )^{2h}}  = \frac{1}{\cosh^{4h}(\alpha t)} \frac{1}{(1-\tanh^2(\alpha t) )^{2h}} = 1 ~. \nn
\end{align}
We have used $\mu^2 + \lambda^2 =h$  and the master identity \cref{master} with the identification $y_p=2h/p$ in the second step. In the third step, we used $|\alpha_-|^2 = \tanh^2(\alpha t)$. 

\subsubsection*{Stress tensor}

The evolved wavefunction for the stress tensor was found to be
\begin{align}\label{stress-state}
	\Psi_T (u)= - \sqrt{2\over c}e^{2\alpha_0}  \sum_{m=2}^{\infty} (\alpha_-)^{m-2} \left[ \sum_{j=1}^{m-1} j(m-j)u_ju_{m-j} - 2m(\mu m-i\lambda)u_m\right]~. 
\end{align}
We have the condition $\mu=+i\lambda$ enforcing $h=0$ as the stress tensor is a descendant of the vacuum; we need to set $\mu=-i\lambda$ for the conjugate wavefunction while taking inner products. These complications can be avoided by setting $\lambda^2=-\mu^2=-\frac{c-1}{24}$ at the very end of the calculation. 
These monomials in \eqref{stress-state} above are not normalized to unity. We have the following cases for non-vanishing inner products.  
\begin{align}\label{uu-norms}
	(u_ju_{m-j},u_pu_{m-p}) \equiv {\cal M}_{j,p}
	=\begin{cases}
		\frac{1}{4j(m-j)} \quad \text{if }j=p \\
		\frac{1}{4j(m-j)} \quad \text{if }m-j=p \\ 
		\frac{2}{m^2}\qquad\quad  \!\text{if }j=p=m/2 \text{ for even } m
	\end{cases}\hspace{-0.5cm},~ (u_m,u_m)  = \frac{1}{2m}\,.
\end{align}
Let us verify that the norm of \eqref{stress-state} is unity. The evaluation turns out to be rather technically subtle and the calculation of the K-complexity \eqref{ko-1} involve similar manipulations. Therefore, we go though this in some of amount of detail
\begin{align}\label{TT-norm}
	\big( \Psi_T (u), \Psi_T (u)  \big)=&\frac{2}{c}e^{4\alpha_0}  \sum_{m=2}^{\infty} |\alpha_-|^{2m-4} \left[  \sum_{j,p=1}^{m-1}  j^2(m-j)^2 {\cal M}_{j,p}  + {2m}(\mu^2 m^2{+\lambda^2}) \right] ~. 
\end{align}
We can now set $\lambda^2=-\mu^2$. 
The sum over $j,\,p$ needs some care when $m$ is even due to last line of \cref{uu-norms}. We shall therefore consider odd and even $m$ cases separately.   For $m$  odd we have
\begin{align}\label{m-odd-sum}
\! \sum_{j,p=1}^{m-1}  j^2(m-j)^2 {\cal M}_{j,p}  + {2m}(\mu^2 m^2-\mu^2 ) 
	= \frac{1}{2}  \sum_{j=1}^{m-1}   j(m-j) + \frac{c-1}{12} m(m^2 - 1)   
	=  
	    \frac{c}{12} m(m^2 - 1). 
\end{align}
The even $m$ case also gives the same result, but the mechanism is different as we need to treat the $j=p=m/2$ term separately -- see \cref{uu-norms}. 
\begin{align}\label{m-even-sum}
	&	 \sum_{j,p=1}^{m-1}  j^2(m-j)^2 {\cal M}_{j,p}  + {2m}(\mu^2 m^2-\mu^2 ) \cr
	=~&  \frac{1}{2} \left[\sum_{j=1}^{m/2-1} + \sum_{j=m/2+1}^{m-1} \right]   j(m-j) + \frac{m^2}{8}+ \frac{c-1}{12} m(m^2 - 1)  
	=  \frac{c}{12} m(m^2 - 1) ~. 
\end{align}
The $m^2/8$ term is the contribution from $j=p=m/2$. 

Returning to the norm \eqref{TT-norm}, we then have
\begin{align}\label{TT-norm-2}
	\big( \Psi_T (u), \Psi_T (u)  \big)=~&\frac{2}{c}e^{4\alpha_0}  \sum_{m=2}^{\infty} |\alpha_-|^{2m-4} \left[ \frac{c}{12} m(m^2 - 1)  \right]\nn \\  =~& e^{4\alpha_0} \frac{1}{(1-|\alpha_{-}|^2)^4} 
	= \frac{1}{\cosh^8(\alpha t)} \frac{1}{(1-\tanh^2(\alpha t))^4}  = 1~, 
\end{align}
where, we used $e^{\alpha_0h}=\text{sech}^{2h}(\alpha t)$ and $|\alpha_-|^2=\tanh^2(\alpha t)$. 
\begin{small}
\providecommand{\href}[2]{#2}
\end{small}
\end{document}